\documentclass[preprint,floats,tightenlines,11pt,eqsecnum,showpacs,aps]{revtex4}
\usepackage{amsmath}
\usepackage{graphicx}
\begin{document}
\def\appls{\hbox{$<$\kern-.75em\lower 1.00ex\hbox{$\sim$}}}

%\title{CYCLIC COSMOLOGY WITH anti de-SITTER SPACETIME}

\title{WAVE EQUATION FOR THE SCALE FACTOR IN A CYCLIC UNIVERSE\\ AND THE ACCELERATION-DECELERATION TRANSITIONS}

\title{SERIAL ACCELERATION-DECELERATION TRANSITIONS\\IN A CYCLIC UNIVERSE
WITH anti de-SITTER SPACETIME}

\title{ACCELERATION-DECELERATION TRANSITIONS\\ IN A CYCLIC UNIVERSE
WITH anti de-SITTER SPACETIME}

\title{SERIAL ACCELERATION-DECELERATION TRANSITIONS\\ IN A CYCLIC UNIVERSE WITH NEGATIVE CURVATURE}

\author{Miloslav Svec\footnote{svec@hep.physics.mcgill.ca}}
%\affiliation{Physics Department, Dawson College, Montreal, Quebec, Canada H3Z 1A4}
\affiliation{Physics Department, Dawson College, Westmount, Quebec, Canada H3Z 1A4}

%\date{November 02, 2017}
%\date{March 5, 2018}
%\date{April 5, 2018}
%\date{May 29, 2018}
%\date{August 31, 2018}
%\date{September 17, 2018}
%\date{October 16, 2018}
%\date{October 23, 2018}
\date{November 20, 2018}

\begin{abstract}

In this work we develop a general phenomenological model of the Cyclic Universe. We construct periodic scale factor $a(t)$ from the requirement of the periodicity of $a(t)$ with no singular behaviour at the turning points $t_\alpha$ and $t_\omega$ and the requirement that a unique analytical form of the Hubble function $H(z)$ can be derived from the Hubble function $H(t)$ to fit the data on $H(z)$. We obtain two versions of $a(t)$ called Model A and Model C. Hubble data $H(z)$ select Model A. With the analytical forms of the Hubble functions $H(t)$ and $H(z)$ known we calculate the deceleration parameters $q(t)$ and $q(z)$ to study the acceleration-deceleration transitions during the expansion phase. We find that the initial acceleration at $t_\alpha=0$ transits at $t_{ad1}=3.313\times10^{-38}$s into a deceleration period that transits at $t_{da}=6.713$Gyr to the present period of acceleration. The present acceleration shall end in a transition to the final deceleration at $t_{ad2}=38.140$Gyr. The expansion period lasts 60.586Gyr. The complete cycle period is $T=121.172$Gyr. 

We use the deceleration parameters $q(z)$ and $q(t)$ to solve the Friedmann equations for the energy densities of Dark Energy $\Omega_0$ and Dark Matter $\Omega_M$ to describe their evolutions over a large range of $z$ and $t$. We show that in Model A the curvature density $\Omega_c(z)$ evolves from a flat Universe in the early times to a curved anti de-Sitter spacetime today. There is no Standard Model Inflation in the Model A.

In the Model A the entire evolution of the Cyclic Universe is described by an energy function - the Hubble function $H(t)$. The evolution proceeds by the minimization of this energy function from $+H_{max}>0$ to $-H_{max}<0$ at times $t^*$=2.175x10$^{-38}$s$>0$ and $T-t^*$, respectively, followed by a rapid phase transition from $-H_{max}<0$ to $+H_{max}>0$. There is no Big Bang singularity at $t_\alpha$ or $t_{2\alpha}=T$ where $H(t_\alpha)=H(t_{2\alpha})=0$. The dynamics of energy function minimization may represent a general evolution principle of the Universe.

\end{abstract}
\pacs{9880.-k, 9535.+d, 9535.+x}

\maketitle

\tableofcontents

\newpage
\section{Introduction}

Over the past century our view of the Universe has been evolving from seeing the Universe as a static and stable system to imagining it expanding at a constant velocity to assuming that this velocity is decreasing at a constant rate to recent observations that this velocity is actually accelerating~\cite{perlmutter97,perlmutter99,riess98,schmidt98}. These observations are embodied in the Hubble function $H^2(z)$ of a highly successful $\Lambda$CDM Model~\cite{planck15,scott18} as a constant Dark Energy density term $\rho_\Lambda$ with equation of state $w_\Lambda=-1$ corresponding to a negative pressure $p_\Lambda=w_\Lambda\rho_\Lambda$. The physical origin of the Dark Energy is unknown in the $\Lambda$CDM Model and in the numerous alternative models of Dark Energy~\cite{amendola10,joyce16,brax18}.

In the $\Lambda$CDM Model and in some other cosmological models the evolution of the Universe begins with an initial singularity~\cite{hawking70,hawking96} with the scale factor $a(t_i)=0$ at the initial time $t_i=0$ corresponding to the redshift $z_i=\infty$ where $H^2(z_i)=\infty$. The fate of the Universe is unknown as it depends on the unknown future evolution of Dark Energy with some models predicting a catastrophic breakdown of the spacetime itself. An appealing solution is a slowdown of the expansion leading to a contraction of the Universe. To avoid the initial singularity various Bouncing Cosmologies have been proposed (for recent reviews see Ref.~\cite{novello08,brandenberger17}). An effective realization of the non-singular cosmological models are the Cyclic Models of the Universe, such as the recent model of the Ekpyrotic Universe~\cite{khoury01,buchbinder07}. 

In this paper we develop a general phenomenological model of a Cyclic Universe. We start with the assumption that the scaling factor $a(t)$ is a finite periodic function with a non-zero minimum $a_{min}>0$ at the initial turning point $t_\alpha$ of the expansion of the Universe and a finite maximum $a_{max}>0$ at the final turning point $t_\omega$ of the expamsion followed by a return to $a_{min}$ during the contraction of the Universe. This means that at the turning points the Hubble function $H(t)=\frac{1}{a}\frac{da}{dt}$ is non-singular with $H(t_\alpha)=H(t_\omega)=0$ in a radical departure from the $\Lambda$CDM Model.

Our task is to determine the analytical forms for the scale factor $a(t)$, the time dependent Hubble parameter $H(t)$ and the redshift dependent parameter $H(z)$. At our disposal is the assumption of the cyclicity embodied in the general mathematical properties of the periodic functions $a(t)$ and $H(t)$ and the requirement that a unique analytical form of $H(z)$ can be derived from $H(t)$ to fit the data on $H(z)$. We call this powerfull requirement "solvability". It is the requirement of "solvability" that constraits the analytical form of the scale factor and Hubble function to simple forms, both as a function of the cosmic time $t$ and the redshift $z$. We present two such  forms called Model A and Model C. The fits to the Hubble data $H(z)$ select the Model A.

With the analytical forms of the Hubble functions $H(t)$ and $H(z)$ we can calculate the deceleration parameter $q(t)$ and $q(z)$ to study the acceleration-deceleration transitions during the expansion phase. We find that the initial acceleration at $t_\alpha$ transits at $t_{ad1}=3.313\times10^{-38}$s into a deceleration period that transits at $t_{da}=6.713$Gyr to the present period of acceleration. The present acceleration shall end in a transition to the final deceleration at $t_{ad2}=38.140$Gyr. The expansion period lasts 60.586Gyr. The complete cycle period is $T=121.172$Gyr. There is no Standard Model Inflation in the Model A. 

The paper is organized as follows. In Section II we introduce Friedmann equations for the Cyclic Universe and discuss its general features. In Section III we present a simple analytical model of Cyclic Universe that inspired this work. In Section IV we present  advanced Models A and C. Their fits to Hubble data in Section V select Model A. We analyse the evolution of the Cyclic Universe near the turning points in Section VI. The serial acceleration-deceleration transitions are described in detail in Section VII. In Section VIII we use the known deceleration parameters $q(z)$ and $q(t)$ to solve the Friedmann equations for the densities of Dark Energy $\Omega_0$ and Dark Matter $\Omega_M$ to describe their evolutions over a large range of $z$ and $t$. In Section IX we use the present curvature density parameter $\Omega_{c,0}>0$ predicted in the sequel paper~\cite{svec17e} and measured in~\cite{svec17b} to show that in Model A the curvature density $\Omega_c(z)$ evolves from a flat Universe in the early times to a curved anti de-Sitter spacetime today. The paper closes with a summary in the Section X and Appendix.

%\newpage
\section{The Cyclic Universe.}

\subsection{Friedmann equations for the Cyclic Universe}

We assume a homogeneous and isotropic spacetime with Robertson-Walker (RW) metric. In cartesian coordinates it is given by~\cite{weinberg08,carroll04}
\begin{eqnarray}
g_{ij} & = & a(t)^2(t)\Bigl(\delta_{ij}+\frac{k}{R_0^2}\frac{x^ix^j}{1-\frac{k}{R_0^2}\vec{x}^2} \Bigr)\\
\nonumber
g_{i0} & = & 0, g_{00}=-1
\end{eqnarray}
where $R_0$ is the curvature parameter and $k=-1,0,+1$ stands for open, flat and closed geometry. For a homogeneous and isotropic cosmic fluid with energy density $\rho$ and pressure $p$ Friedmann equations have the form
\begin{eqnarray}
\rho+\rho_\Lambda+\rho_c & = & \frac{3c^2}{8\pi G} H^2\\
p+p_\Lambda+p_c & = & \frac{3c^2}{8\pi G}\bigl(-H^2-\frac{2}{3}\frac{dH}{dt} \bigr)
\end{eqnarray}
Here $\rho_\Lambda$ and $p_\Lambda=-\rho_\Lambda$ are the energy density and pressure of cosmological constant and $\rho_c$ and 
$p_c=-\frac{1}{3}\rho_c$ are the energy density and pressure of the curvature~\cite{carroll04}. These two energy densities are given by
\begin{equation}
\rho_\Lambda = \frac{3c^2}{8\pi G} \frac{\Lambda}{3}, \quad
\rho_c = \frac{3c^2}{8\pi G}\frac{-kc^2}{R_0^2a^2}
\end{equation}
where $\Lambda$ is the cosmological constant. They satisfy continuity equations
\begin{eqnarray}
\frac{d\rho_\Lambda}{dt} +3H\rho_\Lambda & = & -3Hp_\Lambda\\
\nonumber
\frac{d\rho_c}{dt} +3H\rho_c & = & -3Hp_c
\end{eqnarray}
Using these relations Friedmann equations lead to similar continuity equation for the density $\rho$. The Hubble function is defined in terms of the scale factor
\begin{equation}
H(t)=\frac{1}{a(t)}\frac{da(t)}{dt}
\end{equation}
The scale factor is cyclic with a period $T$ so that $a(t+T)=a(t)$. During the expansion phase $H(t)>0$, during the contraction $H(t)<0$. At the turning points $t_\alpha=0$ and $t_\omega=T/2$ of the expanding  Universe the scale factor $a(t_\alpha)=a_{min}> 0$ and $a(t_\omega)=a_{max}< \infty$. Consequently
\begin{equation}
H(t_\alpha)=H(t_\omega)=0
\end{equation}
The contraction phase ends at the turning point $t_{2\alpha}=T$ with the scale factor $a(t_{2\alpha})=a_{min}$ and $H(t_{2\alpha})=0$. 

Since $H(t)$ is a cyclic function the combinations
\begin{eqnarray}
\bar{\rho}  & = & \rho + \rho_\Lambda + \rho_c\\
\nonumber
\bar{p} & = & p + p_\Lambda + p_c
\end{eqnarray}
are the cyclic energy density and the cyclic pressure. The Friedmann equations for the Cyclic Universe then read
\begin{eqnarray}
\bar{\rho} & = & \frac{3c^2}{8\pi G} H^2\\
\bar{p} & = & \frac{3c^2}{8\pi G}\bigl(-H^2-\frac{2}{3}\frac{dH}{dt} \bigr)
\end{eqnarray}
where $\bar{\rho}$ and $\bar{p}$ satisfy continuity equation 
\begin{equation}
\frac{d\bar{\rho}}{dt} +3H\bar{\rho} = -3H\bar{p}
\end{equation}
Notice that $H(t)$ does not depend on the curvature parameter $R_0$ and therefore on $\rho_c$, $p_c$.

\subsection{General features of the Cyclic Universe}

Inverting the relation (2.6) we find
\begin{equation}
\frac{a(t)}{a(t_\alpha)}=\exp\Bigl[\int \limits_{t_\alpha}^t H(t')dt' \Bigr]
\end{equation}
We shall identify the fixed comoving radius of the Universe with the curvature parameter $R_0$. Then the proper radius of the Universe at time $t$ is given by~\cite{carroll04} 
\begin{equation}
R(t)=R_0a(t)
\end{equation}
The equation (2.12) relates the proper radius and the proper volume $V(t)=\frac{4\pi}{3}R^3(t)$ to the Hubble parameter
\begin{eqnarray}
R(t) & = & R_\alpha \exp\Bigl[\int \limits_{t_\alpha}^t H(t')dt' \Bigr]\\
V(t) & = & V_\alpha \exp\Bigl[3 \int \limits_{t_\alpha}^t H(t')dt' \Bigr]
\end{eqnarray}
where $R_\alpha$ and $V_\alpha$ are the initial proper radius and initial proper volume of the Universe, respectively. 

We shall assume that $H(t)$ and its derivative $\frac{dH}{dt}$ are continous functions of time $t$. We expect that during the early phase of the expansion the Hubble parameter is rapidly increasing (inflation) from zero at $t_\alpha$ reaching a maximum $H_{max}>0$ and then monotonically decreasing until a final decrease (pre-deflation) near $t_\omega$ to zero at the turnig point $t_\omega$. Following the early phase of the contraction (deflation) the negative Hubble parameter continues to monotonically decrease reaching a minimum $H_{min}<0$, and then rapidly  increasing (pre-inflation) near $t_{2\alpha}$ to zero at the turnig point $t_{2\alpha}$. This behaviour of the Hubble parameter implies a period of rapid inflation of the proper volume of the Universe during the initial stage of its expansion, and a period of rapid deflation of the proper volume of the Universe during the final stage of its contraction (pre-inflation).

At the turning points $\bar{\rho}(t_\alpha)=\bar{\rho}(t_\omega)=0$ and there are finite pressures 
\begin{eqnarray}
\bar{p}(t_\alpha) & = & -\frac{c^2}{4\pi G}\frac{dH(t_\alpha)}{dt}<0\\
\nonumber
\bar{p}(t_\omega) & = & -\frac{c^2}{4\pi G}\frac{dH(t_\omega)}{dt}>0
\end{eqnarray}
There are no spatial or spatio-temporal singularities at the turning points, and no breakdown of the Einstein theory of gravity. The Energy Conditions~\cite{carroll04} for singular universes with $a_{min}=0$ and $a_{max}=\infty$ cannot exclude non-singular cyclic universes. 

It is useful to introduce the deceleration parameter
\begin{equation}
q(t)=\frac{-1}{H^2a}\frac{d^2a}{dt^2}=-1-\frac{1}{H^2}\frac{dH}{dt}=-1+\frac{d}{dt}\frac{1}{H}
\end{equation}
Then the expression for the cyclic pressure takes the form
\begin{equation}
\bar{p}=\bar{\rho}\Bigl[-\frac{1}{3}+\frac{2}{3}q(t) \Bigr]
\end{equation}
For $q=\frac{1}{2}$ we get $\bar{p}=0$ which is the point where the pressure changes sign. In principle there could be more than one such points. At the turning points deceleration parameter diverges
\begin{eqnarray}
q(t \rightarrow t_\alpha^-) & = & 
q(t \rightarrow t_\alpha^+) = -\infty\\
\nonumber
q(t \rightarrow t_\omega^-) & = & 
q(t \rightarrow t_\omega^+) = +\infty\\
\end{eqnarray}
This is a classical $\lambda$ type phase transition akin e.g. to $\lambda$ phase transition of specific heat~\cite{greiner94}. It is called $"\lambda"$ since the divergent behaviour of the order parameter resembles the letter $\Lambda$. In our case the order parameter is the deceleration parameter which describes the periodic phase transitions between two phases of the evolution of the Universe: the expansion and the contraction. The divergent behaviour of this order parameter should not be confused with spatial singularities.

The homogeneity and isotropy of the spacetime implies a fundamental relation of the scale factor with the redshift~\cite{ryden03}
\begin{equation}
1+z=\frac{a(t_0)}{a(t)}
\end{equation}
where $t_0$ is the present time. The Planck spectrum of the Cosmic Microwave Background (CMB) implies additional relation for the temperature of the Universe~\cite{serjeant10}
\begin{equation}
\frac{T(t)}{T_0}=\frac{a(t_0)}{a(t)}
\end{equation}
where $T_0=2.7255$ K is the present temperature of the Universe. Then $T(t)=T_0(1+z)$. We identify the initial temperature $T(t_\alpha)$ with the Planck temperature $T_{Pl}=1.417 \times 10^{32}$ K which defines the finite initial redshift $z_\alpha$
\begin{equation}
1+z_\alpha=\frac{T_{Pl}}{T_0}=\frac{a(t_0)}{a(t_\alpha)}
\end{equation}

\section{Simple cyclic model.}

We seek a periodic Hubble parameter $H(t+T)=H(t)$ such that $H(0)=H(T/2)=H(T)=0$, $H(t)>0$ for $0<t<T/2$ and $H(t)<0$ for $T/2 <t <T$. Since $H(t)$ is a periodic function it can be expressed as a time series
\begin{equation}
H(t)=H(0)+\sum \limits_{n=1}^{\infty} A_n\sin(n\Omega t)+B_n \cos(n\Omega t)
\end{equation}
Since $H(-t)=-H(t)$ we have $H(0)=B_n=0$. Then
\begin{equation}
H(t)=\sum \limits_{n=1}^{\infty} A_n\sin(n\Omega t)
\end{equation}
and
\begin{equation}
\int H(t')dt'=- \sum \limits_{n=1}^\infty \frac{A_n}{\Omega}\frac{\cos(n\Omega t)}{n}
\end{equation}
We require that both sums are summable to give an analytical expression. A list of Fourier series in Gradshteyn and Ryzhik Tables~\cite{gradshteyn80} lists one such pair of series
\begin{eqnarray}
\sum \limits_{n=1}^\infty p^n \sin(nx) & = & \frac{p\sin x}{1-2p\cos x + p^2}, \quad p^2 <1\\
\sum \limits_{n=1}^\infty p^n \frac{\cos(nx)}{n} & = & \ln \frac{1}{\sqrt{1-2p\cos x + p^2}}, \quad p^2 < 1
\end{eqnarray}
With $x\equiv \phi =\Omega t$ we define Hubble parameter
\begin{equation}
H(t)=\frac{\Omega p \sin \phi}{1-2p\cos \phi + p^2} = \frac{\Omega p \sin \phi}{F(\cos \phi)}
\end{equation}
and the scale factor
\begin{equation}
a(t)=a(t_0)\exp\Bigl[\int \limits_{t_0}^t H(t')dt' \Bigr]= a(t_0)\frac{\sqrt{F(\cos \phi)}}{\sqrt{F( \cos \phi_0)}}
\end{equation}
where we have defined $F(\cos \phi)=1-2p\cos \phi + p^2$ and where $\phi_0=\Omega t_0$. Then the redshift relation (2.21) reads
\begin{equation}
1+z=\frac{\sqrt{F(\cos \phi_0)}}{\sqrt{F(\cos \phi)}}
\end{equation}
Since $p^2<1$ we can set $p=1-\delta$. At $t=0$ the phase $\phi=0$ and $F(1)=(1-p)^2=\delta^2$ so that (2.23) reads
\begin{equation}
1+z_\alpha=\frac{T_{Pl}}{T_0}=\frac{\sqrt{F(\cos \phi_0)}}{\delta}
\end{equation}
We see that $\delta$ is a very small number so that for $z\ll z_\alpha$ we can set $p=1$. For $\phi$ near 0 $F(\cos \phi)=2\delta$ and the Hubble parameter becomes large and increasing $H(t)=\frac{\Omega^2 t}{2\delta}$.

At $t_0$ we can write $H_0=\frac{\phi_0}{t_0}\frac{\sin \phi_0}{F(\cos \phi_0)}$. With Planck 2015 values~\cite{planck15} $H_0=67.81$ kms$^{-1}$Mpc$^{-1}$ and $t_0=13.799$ Gyr we determined from this expression the value $\phi_0=41.00^\circ$ which implies $T=\frac{2 \pi t_0}{\phi_0}=121.162$ Gyr. The expansion (contraction) lasts 60.581 Gyr.

Next we need to determine the analytical form of $H(z)$ to compare the model with the data. The system of equations (3.6) and (3.8) is analytically solvable for $H(z)$. From (3.8) we find 
\begin{equation}
F(\cos \phi)=1+p^2-2p\cos \phi=\frac{F(\cos \phi_0)}{(1+z)^2}
\end{equation}
Substituting into (3.6) we have
\begin{equation}
H(z)=H_0\frac{\sin \phi}{\sin \phi_0} (1+z)^2
\end{equation}
Solving (3.10) for $\cos \phi$ in terms of $z$ we get
\begin{equation}
\cos \phi(z) =\frac{(1+p^2)(1+z)^2-F(\cos \phi_0)}{2p(1+z)^2}
\end{equation}
From here we find $\sin \phi(z)$ and putting it all together we obtain
\begin{equation}
H(z)=H_0\frac{\Bigl[(1+p^2)(1+z)^2-F(\cos\phi_0)\Bigr]^{\frac{1}{2}} 
              \Bigl[-(1-p^2)(1+z)^2+F(\cos\phi_0)\Bigr]^{\frac{1}{2}}}
              {2p\sin \phi_0}
\end{equation}              
For $p=1$ $F(\cos\phi_0)=2(1-\cos \phi_0)$ and
\begin{equation}
H(z)=H_0\frac{\Bigl[(1+z)^2-1+\cos\phi_0\Bigr]^{\frac{1}{2}} 
              \Bigl[1-\cos\phi_0\Bigr]^{\frac{1}{2}}}
              {\sin \phi_0}
\end{equation}
A comparison of $H(z)$ from (3.14) with the Hubble data shown in the Table VI in the Appendix gives an encouraging but still a poor $\chi^2=2.15/dof$. This compares with $\chi^2/dof=0.7680$ for $\Lambda$CDM Model. Introducing an overall normalization factor $B$ improved the $\chi^2$ for $B>1$ but not enough.

Assuming $B=1$ we can calculate the pressure and the deceleration parameter. The expression for the pressure reads
\begin{equation}
\bar{p}=\frac{1}{8\pi G} \Bigl[H^2-\frac{2\Omega^2p\cos \phi}{1-2p\cos \phi +p^2} \Bigr]
\end{equation}
For $\phi=0$ the initial pressure $\bar{p}_\alpha=-\frac{1}{4\pi G}\frac{\Omega^2}{\delta^2} <0$. For $\phi=\pi$ the final pressure $\bar{p}_\omega=+\frac{1}{4\pi G}\frac{\Omega^2}{4}>0$. The condition $\bar{p}=0$ where the pressure is changing signs has a solution for $\cos \phi_p$
\begin{equation}
\cos \phi_{p(1,2)}=\frac{1}{3}\Bigl[2\pm \sqrt{1-p^2+p^4}\Bigr]=\frac{1}{3}\bigl[2\pm(1+\delta)\bigr]
\end{equation}
The solution $\cos \phi_{p,1}=1+\frac{1}{3}\delta >1$ is unphysical. The solution $\cos \phi_{p,2}=\frac{1}{3}(1-\delta)=\frac{1}{3}$ yields $\phi_{p,2}=70.53^\circ$. With $\Omega = \frac{\phi_0}{t_0}=0.051858$ rad Gyr$^{-1}$ we find the time the pressure changes sign $t_p=\frac{\phi_{p,2}}{\Omega}=23.737$ Gyr.

The deceleration parameter is given by
\begin{equation}
q(t)=\frac{-p\cos^2\phi-(1+p^2)\cos\phi +2p}{p\sin^2\phi}
\end{equation}
Evidently $q_\alpha=-\infty$ and $q_\omega=+\infty$. Solving $q=0$ for $\cos\phi$ yields only one physical solution $\cos \phi_q =p=1-\delta$.  With $\cos \phi_q=1-\frac{1}{2}\phi_q^2$ we find $\phi_q^2 =2\delta$. After the initial acceleration there is a constant deceleration for $\phi>\phi_q$ in contradiction with the observation of the late time acceleration. With a poor $\chi^2/dof$ and no re-acceleration this Simple Model is rejected. However the ideas of this model inspire a general method ("general solvability") to construct improved physical models.

\section{Advanced cyclic models.}

\subsection{Model A}

The construction of a model of cyclic Hubble function $H(t)$ is constrained by the determination of the analytical form of $H(z)$ to compare the cyclic model with the data. This determination is made possible by three assumptions inspired by the Simple Model: (1) $H(t)$ and the redshift $1+z$ (scale factor $a(t)$) both depend explicitely on the function $F(\cos\phi)$ (2) There is a unique analytical solution for $F(\cos \phi)$ in terms of $z$ (3) The term $\sin \phi$ can be calculated from the expression for $\cos \phi$ derived from $F(\cos \phi)$. We call this principle "general solvability".

Since Hubble parameter is odd under time reversal, the scale factor is even. This allows us to assume that the scale factor is a function of $F=F(\cos\phi)$ where $\phi(t)=\Omega t$. Then
\begin{equation}
H(t) = \frac{1}{a}\frac{da}{dt}=\frac{1}{a}\frac{da}{dF}\frac{dF}{d\cos\phi}\frac{d\cos\phi}{dt}=\Omega \sin\phi G(F)
 \end{equation}
where
\begin{equation}
G(F) =  \frac{2p}{a}\frac{da}{dF}
\end{equation}
The equation (4.1) is the most general form of cyclic $H(t)$. Inspired by the behavour of the redshift $1+z$ at $\phi=0$ in the Simple Model we shall assume a general form of the scale factor
\begin{equation}
a(F)=\frac{F^n}{f(F)}
\end{equation}
where $n>0$. Then the redshift
\begin{equation}
1+z=\frac{F_0^n}{f(F_0)}\frac{f(F)}{F^n}
\end{equation}
where $F_0=F(\cos \phi_0)$. We require that there be a unique analytical solution for $F=F(z)$. This seems only possible when we set $f(F)=A-BF^n$ where $A>0$, $B>0$. Then
\begin{equation}
1+z=\frac{F_0^n}{A-BF^n_0}\frac{A-BF^n}{F^n}
\end{equation}
can be solved for $F(z)$
\begin{equation}
F(z)=F_0 \Bigl[\frac{1}{1+(1-CF_0^n)z} \Bigr]^{\frac{1}{n}}=1+p^2-2p \cos \phi
\end{equation}
where $C=B/A$ and $0<1-CF_0^n)z<1$. Solving (4.6) for $\cos \phi$ with $\phi=\Omega t$ we determine time as a unique function of redshift
\begin{equation}
t=t(z)=\frac{1}{\Omega}\arccos \Bigl(\frac{1+p^2-F(z)}{2p} \Bigr)
\end{equation}
With the final definition of the scale factor
\begin{equation}
a(F)=\frac{F^n}{1-CF^n}
\end{equation}
and with $F=F(t)=1+p^2-2p \cos \Omega t$ the Hubble function $H=H(t)$ is then given by
\begin{equation}
H(t)=\Omega \sin\phi \frac{2np}{F(t)(1-CF^n(t))}
\end{equation}
Solving for (4.6) for $\cos \phi$ with $\phi=\phi(z)$ we determine 
\begin{eqnarray}
\sin \phi(z) & = & \frac{1}{2p}\Bigl[(1+p)^2-F(z)\Bigr]^{\frac{1}{2}}
\Bigl[-(1-p)^2+F(z)\Bigr]^{\frac{1}{2}}\\
\nonumber
  & = & \frac{1}{2p}\Bigl[-(1-p^2)+2(1+p^2)F(z)-F^2(z)\Bigr]^{\frac{1}{2}}
\end{eqnarray}
With $F=F(z)$ given by (4.6) the Hubble function $H=H(z)$ then reads for any $p$
\begin{equation}
H(z)=\Omega n \Bigl[-(1-p^2)+2(1+p^2)F(z)-F^2(z)\Bigr]^{\frac{1}{2}}\frac{1}{F(z)(1-CF^n(z))}
\end{equation}
At $z=0$ we find $\Omega n$
\begin{equation}
\Omega n =H_0 \frac{F_0(1-CF^n_0)}{\Bigl[-(1-p^2)+2(1+p^2)F_0-F^2_0\Bigr]^{\frac{1}{2}}}
\end{equation}
Notice that this relation enables us to determine $\Omega$ from $H_0$ and the fitted paramaters $F_0,n,C$. The Hubble function $H(z)$ for any $p$ then reads
\begin{eqnarray}
H(z) & = & H_0\frac{\sin \phi(z)}{\sin \phi_0}
        \frac{F_0(1-CF_0^n)}{F(z)(1-CF^n(z))}\\
\nonumber
     & = & H_0 
     \Biggl\{\frac{-(1-p^2)+2(1+p^2)F(z)-F^2(z)}{-(1-p^2)+2(1+p^2)F_0-F^2_0}\Biggr\}^{\frac{1}{2}} \frac{F_0(1-cF_0^n)}{F(z)(1-CF^n(z))}
\end{eqnarray}
With $p=1$ approximation at $z \ll z_\alpha$ we find
\begin{equation}
H(z)=H_0\frac{1+Dz}{1+z}\Bigl[\frac{4(1+Dz)^{\frac{1}{n}} -F_0}{4-F_0} \Bigr]^{\frac{1}{2}}
\end{equation}
where $D=1-CF_0^n$. The equations (4.9) and (4.14) for $H(t)$ and $H(z)$ constitute our Model A. 

We can also express the Huble function as a function of the scale factor $a$. Inverting (4.8) we get for $F=F(a)$
\begin{equation}
F=\frac{a^\frac{1}{n}}{(1+aC)^\frac{1}{n}}=1+p^2-2p \cos \phi(a)
\end{equation}
Solving for $\sin \phi(a)$ and substituting into (4.9) we find $H=H(a)$
\begin{equation}
H=\frac{n \Omega (1+aC)}{a^\frac{1}{n}}
\sqrt{\Bigl[+a^\frac{1}{n}-(1+p)^2(1+aC)^\frac{1}{n}\Bigr]
      \Bigl[-a^\frac{1}{n}+(1-p)^2(1+aC)^\frac{1}{n}\Bigr]}
\end{equation}
In the $p=1$ approximation far away from $a \ll 1$ this becomes
\begin{equation}
H=\frac{n \Omega (1+aC)}{a^\frac{1}{2n}}
\sqrt{4(1+aC)^\frac{1}{n}-a^\frac{1}{n}}
\end{equation}
For $a \ll 1$ we obtain
\begin{equation}
H=\frac{2n\Omega}{a^\frac{1}{n}}
\sqrt{a^\frac{1}{n}-a^\frac{1}{n}_\alpha}
\end{equation}
where $a_\alpha=a(t=0)=(1-p)^{2n}$.

\subsection{Model C}

The scale factor of the Model A can be generalized 
\begin{equation}
a(F)=\Bigl[\frac{F^n}{1-CF^n}\Bigr]^m
\end{equation}
where $m>0$. The Hubble function $H(t)$ has the form
\begin{equation}
H(t)=\Omega \sin\phi \frac{2mnp}{F(t)(1-CF^n(t))}
\end{equation}
and the redshift reads
\begin{equation}
1+z=\Bigl[\frac{F_0^n}{1-CF^n_0}\Bigr]^m\Bigl[\frac{1-CF^n}{F^n}\Bigr]^m
\end{equation}
Solving for $F(z)$ we get
\begin{equation}
F(z)=\frac{F_0}
{\Bigl[(1+z)^{\frac{1}{m}}(1-CF_0^n)+CF_0^n\Bigr]^{\frac{1}{n}}}=1+p^2-2p \cos \phi
\end{equation}
With this $F(z)$ we can again calculate $t=t(z)$ and $\sin \phi(z)$. For any $p$ the Hubble function $H(z)$ has still the form of (4.13). For $p=1$ at $z \ll z_\alpha$ it reads
\begin{equation}
H(z)=H_0\frac{1+D\big[(1+z)^{\frac{1}{m}}-1\bigr]}{(1+z)^{\frac{1}{m}}}
\frac{\Bigl[4\bigl[1+D\bigl((1+z)^{\frac{1}{m}}-1\bigr)\bigr]^{\frac{1}{n}}-F_0\Bigr]^{\frac{1}{2}}}{\bigl(4-F_0)^{\frac{1}{2}}}
\end{equation}
For $m=1$ we recover the form (4.14). We shall refer to the Hubble functions (4.20) and (4.23) as Model C.

\subsection{Deceleration parameter $q_0$ in the Models A and C}

Parameters $F_0$, $C$, $n$ and $m$ are free parameters of the Model C. Model A has fixed $m=1$. Except for $F_0$ these parameters do not appear to have a clear physical interpretation. An important prediction of the $\Lambda$CDM Model is the value of present deceleration parameter $q_0=-0.538$. To facilitate comparisons of our Models A and C with the 
$\Lambda$CDM Model we replace the parameter $C$ with the physically meaningful parameter $q_0$.

The deceleration parameter $q(z)$ is given by 
\begin{equation}
q(z)=-1+\frac{1+z}{H(z)}\frac{dH}{dz}
\end{equation}
We use the expression (4.21) for the Hubble function $H(z)$ in the Model C to calculate $q(z)$. To present the result we first define some useful notations. We define $M=1/m$, $N=1/n$, $Q=(1+z)^M-1$ and
\begin{equation}
R=\frac{4\bigl[1+DQ\bigr]^N-F_0}{4-F_0}
\end{equation}
Then the expression for the deceleration parameter reads
\begin{equation}
q(z)=-1+\frac{H_0M\sqrt{R}}{H}\Bigl[ \frac{D-1}{1+Q} + \frac{1}{R} \frac{2ND\bigl(1+DQ\bigr)^N}{4-F_0}\Bigl]
\end{equation}
Setting $z=0$ we find from (4.24)
\begin{equation}
D=1-CF_0^n=\frac{1+M+q_0}{M}\frac{4-F_0}{4-F_0+2N}
\end{equation}
To obtain these results in the Model A we set $M=1$ so that
\begin{equation}
D = 1-CF_0^n=(2+q_0)\frac{4-F_0}{4-F_0+2N}
\end{equation}
Then with
\begin{equation}
R=\frac{4\bigl[1+Dz\bigr]^N-F_0}{4-F_0}
\end{equation}
we have in the Model A deceleration parameter 
\begin{equation}
q(z) =  -1+\frac{H_0\sqrt{R}}{H}\Bigl[\frac{D-1}{1+z} + \frac{1}{R} \frac{2ND\bigl(1+Dz\bigr)^N}{4-F_0}\Bigl]
\end{equation}
 
\section{Hubble data analysis, results and predictions.}

The most recent measured values of the Hubble parameter used in our fits are from Ref.~\cite{planck15} and Ref.~\cite{jimenez03} - Ref.~\cite{moresco16}. The data are presented in the Table VI in the Appendix. The Table VI also lists the fitted values and errors of the Hubble function of the best fit of the  Model A. We shall refer to these as Hubble data $AH(z)$. The remarkable feature of this data are the very small errors. The input value of $H_0$ was fixed at $H_0=67.81$ kms$^{-1}$Mpc$^{-1}$~\cite{planck15}. We recover this value from the predictions for $H(z=0)$ from all fitted models of $H(z)$.

In actual fits to the Hubble data we fit 4 variants of the cyclic models: Models A and C with $q_0$ a free parameter, and Models B and D which are Models A and C with $q_0$ fixed at the value $q_0=-0.538$ predicted by the $\Lambda$CDM Model. In all runs the initial value of $q_0$ in Models A and C was $q_0=-0.538$. In all runs the initial value of $n$ was the Simple Model value $n=0.5$. 

In the first Run 01 the initial value of the lower limit of $F_0$ was the value $F_0=0.49058$ corresponding to $\phi_0=41.00^\circ$ of the Simple Model. The fitted value of $F_0$ was at this lower boundary. In the runs Run 02 - Run 05 the lower bound on $F_0$ was decreased to values 0.4, 0.3, 0.2 and 0.16175 with the fitted $F_0$ always equal to the lower boundary. For the value $F_0=0.16175$ we get $1-CF_0^n$ very near zero for Models A and C and negative value for Models B and D, which yield unphysical values of the Hubble parameter. In the Run 08 of the Model A we found the value $F_0=0.205233$ to be the limiting value of $F_0$ for which $\frac{dH}{dt}$ is still nonnegative at all $t<T/2$. 

To select the best model we compared the four models in all five runs using the values of $\chi^2/dof=\chi^2_{min}/(N-k)$ where $N$ is the number of data points and $k$ the number of fitted parameters, and the confidence level CL$\%=\exp(-\frac{1}{2}\chi^2/dof)100.0$. In addition we calculated the values of the Akaike and Baysian information criteria, AIC and BIC. These are defined as~\cite{akaike74,schwartz78,burnham02}
\begin{eqnarray}
\text{AIC} & = & \chi^2_{min} +\frac{2kN}{N-k-1}\\
\nonumber
\text{BIC} & = & \chi^2_{min} + k \ln N
\end{eqnarray}
The larger is the difference with respect to the model that carries smaller value of AIC (BIC), the higher is the evidence against the model with larger value of AIC (BIC). 

The results of the Run 01 for all four models and of the Run 08 for the Model A are presented in the Table I and compared with $\Lambda$CDM Model. In Run 01 the best model is the Model A. We exclude the Model C because with more free parameters it has higher values of all selection criteria. The Models B and D have poor values of all selection criteria and are rejected. The values of all selection criteria slightly improve with decreasing $F_0$ in Runs 02-05 which is illustrated in the Table I by the results of Run 08 for the Model A. However the relative merits of the models do not change: in all runs the Model A is the best Model. Comparison of the Model A with the $\Lambda$CDM Model shows convincingly that the Model A fits Hubble data much better than the  $\Lambda$CDM Model.

The Table II presents the results of the fitted and predicted parameters for Models A.01 and A.08. The only large difference between the two models is the value of $F_0$ which leads to large diference of the corresponding value of $\phi_0=\arccos(1-\frac{F_0}{2})$ and therefore large differences in the angular frequency $\Omega=\frac{\phi_0}{t_0}$ and the time period $T=\frac{2\pi}{\Omega}$. In the following we shall work with the Model A.01 and refer to it simply as the Model A.

In Figure 1 we compare Model A and $\Lambda$CDM Model with the Hubble parameter data. The most notable disagreement of the $\Lambda$CDM Model with the data occurs at large $z$ where this model misses completely the value from the BAO measurements at $z=2.34$~\cite{delubac15} while the Model A just passes through this data point.

\begin{table}
\caption{Values of $\chi^2/dof$, information criteria AIC and BIC and the confidence level CL$\%$ for Models A.01, C.01, B.01, D.01 and A.08 compared to the $\Lambda$CDM Model.}
\begin{tabular}{|c||c|c|c|c|}
\toprule 
Model & $\chi^2/dof$ & AIC & BIC &  CL$\%$ \\
\colrule
$\Lambda$CDM & 0.7680 & 26.9269 & 30.0688 & 68.1145 \\
A.01 & 0.5205 & 20.4921 & 23.6340 & 77.0873 \\
C.01 & 0.5417 & 23.2084 & 27.0110 & 76.2742 \\
B.01 & 0.7998 & 26.0568 & 28.3298 & 67.0040 \\
D.01 & 0.7965 & 27.6685 & 30.8104 & 67.1501 \\
A.08 & 0.5173 & 20.4092 & 23.5516 & 77.2102 \\
\botrule
\end{tabular}
\label{Table I.}
\end{table}
\begin{table}
\caption{Values of the fitted parameters $F_0$, $q_0$, $n$ and the predicted parameters $C$, angular frequency $\Omega$, period $T$ and the present time $t_0$ for the Models A.01 and A.08.}
\begin{tabular}{|c||c|c|c|c|c|c|c|}
\toprule
Model & $F_0$ & $q_0$ & $n$ & $C$ & $\Omega$ (rad/Gyr) & $T$ (Gyr) & $t_0$ (Gyr) \\
\colrule
A.01 & 0.49058$\pm$3.31912 & -0.2912$\pm$0.0928 & 0.284589$\pm$0.062417 & 0.527679 & 0.0518535 & 121.172 & 13.800\\
A.08 & 0.20523$\pm$4.34096 & -0.3070$\pm$0.0951 & 
0.302705$\pm$0.062230 & 0.611755 & 0.0329072 & 190.937 & 13.887 \\ 
\botrule
\end{tabular}
\label{Table II.}
\end{table}

\begin{figure} [htp]
\includegraphics[width=12cm,height=10.5cm]{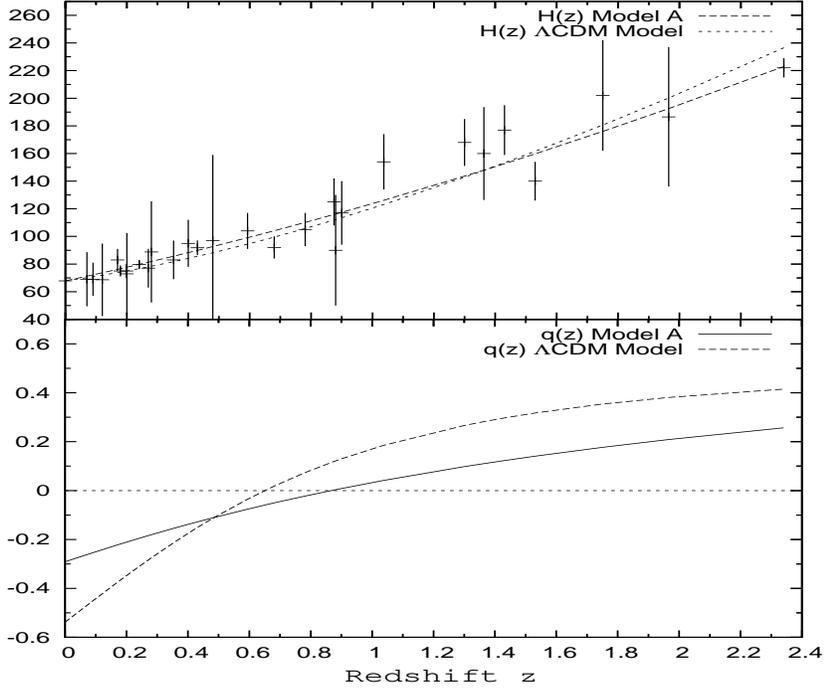}
\caption{Top panel: Fitted Hubble function of the Model A and $\Lambda$CDM Model compared to the Hubble data $H(z)$ in kms$^{-1}$Mpc$^{-1}$. Bottom panel: Predictions for the deceleration parameter $q(z)$ from the Model A and the $\Lambda$CDM Model.}
\label{Figure 1}
\end{figure}

\section{The evolution of the Cyclic Universe in the Model A.}

\subsection{The contraction-expansion period near the initial turning point $t_\alpha=0$}

At the turning point $t_\alpha=0$ the Hubble function $H(t_\alpha)=0$ but it rapidly increases to $H_{max}>0$ in a period of expansion at $t>t_\alpha$ which follows a contrstion period at $t<t_\alpha$ of a rapid increase of $H$ from $H_{min}<0$ to $H(t_\alpha)=0$. To show this behaviour of the Model A we shall work in a small $\phi$ approximation where $\phi=\Omega t$. 

The function $F(\phi)=1+p^2-2p\cos \phi=\delta^2+\phi^2$ where $\delta=1-p$. Then the scale factor $a(F)=\frac{F^n}{1-CF^n}=(\delta^2+\phi^2)^n$. We determine the parameter $\delta$ in terms of the Planck temperature $T_{Pl}$ from the relations (2.21) and (2.22) at $\phi=0$
\begin{equation}
1+z_\alpha=\frac{a(t_0)}{a(t_\alpha)}=
\frac{T(t_\alpha)}{T(t_0)}=\frac{T_{Pl}}{T(t_0)}
\end{equation}
With $a(t_\alpha)=\delta^{2n}$, $T_{Pl}=1.417$x10$^{32}$ K and $a(t_0)=1.434732$, $T_0=2.7255$ K we find $\delta=3.573591$x10$^{-56}$. The physics at this turning point is characterized by the Planck energy $E_{Pl}=1.2210$x10$^{16}$ TeV. 

At small $\phi$ the Hubble function $H(\phi)$ in (4.9) takes the form
\begin{equation}
H(\phi)=\Omega 2 n \frac{\phi}{\delta^2+\phi^2}
\end{equation}
The derivative $\frac{dH}{dt}$ then reads
\begin{equation}
\frac{dH}{dt}=\frac{\Omega^2 2n}{(\delta^2+\phi^2)^2} \Bigl(\delta^2-\phi^2\Bigr)
\end{equation}
There are two solutions for $\frac{dH}{dt}=0$. At $\phi=\delta$ the Hubble function has a maximum $H_{max}=\frac{\Omega n}{\delta}>0$. At $\phi=-\delta$ the Hubble function has a minimum $H_{min}=-\frac{\Omega n}{\delta}<0$. For $-\delta < \phi <+\delta$  the derivative $\frac{dH}{dt}>0$ is positive and describes the increasing $H(t)$ during the contraction-expansion transition  period. For $\phi<-\delta$ and $\phi > +\delta$ the derivative is negative, indicating a decreasing $H(t)$. At the turning point $\phi=0$ the derivative $\frac{dH}{dt}$ is finite and positive. The Friedmann equation (2.10) then implies a finite negative pressure $\bar{p}(t_\alpha)<0$.

\subsection{The expansion-cotraction transition period near the final turning point $t_\omega=T/2$}

For $\phi$ near 180$^\circ$ we set $\phi=180^\circ -\psi$ where $\psi$ is a small angle $>0$ or $<0$. With $p=1$ and in the small angle approximation  $F=1+p^2+2p\cos \psi=4-\psi^2$. Then the scale factor 
\begin{equation}
a(F)=\frac{(4-\psi^2)^n}{1-C(4-\psi^2)^n}
\end{equation}
We can determine the temperature $T_\omega$ of the Universe at the turning point $t_\omega$ from the relation 
$\frac{a(t_0)}{a(t_\omega)}=\frac{T_\omega}{T_0}$. With $a(t_\omega)=6.834258$ we find $T_\omega=0.57217$ K corresponding to the low energy physics scale $E_\omega=0.49304$x10$^{-6}$ eV.

In the small $\psi$ approximation and with $p=1$ the Hubble function reads
\begin{equation}
H(\psi)=\Omega 2n\frac{\psi}{(4-\psi^2)(1-C(4-\psi^2)^n}\approx \Omega 2n\frac{\psi}{4(1-C4^n)}
\end{equation}
In the leading order the derivative $\frac{dH}{dt}$ is given by the expression
\begin{equation}
\frac{dH}{dt}=-\Omega^2 \frac{2n}{4(1-C4^n)} 
\end{equation}
With negative $\frac{dH}{dt}$ the Hubble function decreases to $H(t_\omega)=0$ at the turning point and then continues to decrease to its contraction minimum $H_{min}=-\frac{\Omega n}{\delta}<0$. Since the derivative $\frac{dH}{dt}$ is finite and negative at the turning point $t_\omega$ it follows from the Friedmann equation (2.10) that the pressure $\bar{p}(t_\omega)$ is finite and positive.

\begin{figure} [htp]
\includegraphics[width=12cm,height=10.5cm]{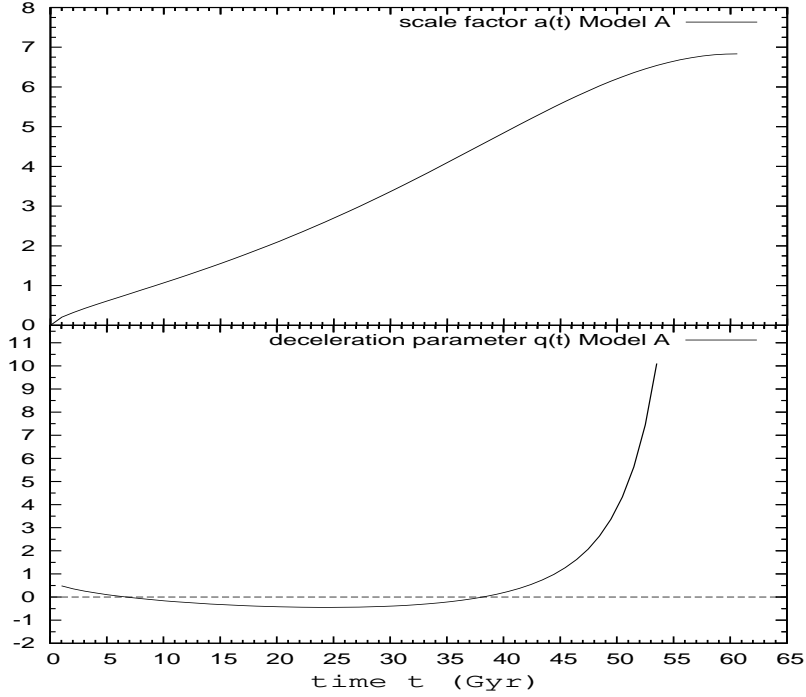}
\caption{Top panel: Evolution of the scale factor $a(t)$ from $t=0.001929$ Gyr to $t_\omega=T/2=60.585$ Gyr. Bottom panel: Evolution of the deceleration parameter $q(t)$ from $t=1.009766$ Gyr to $t=53.517574$ Gyr.}
\label{Figure 2}
\end{figure}

\subsection{The periods of the expansion and contraction}

In general the derivative $\frac{dH}{dt}$ of the Hubble function (4.9) of the Model A is given by the expression
\begin{equation}
\frac{dH}{dt}=H\frac{\Omega \cos \phi}{\sin \phi} -
H^2 \frac{1-(1+n)CF^n}{n}
\end{equation}
With $W=\frac{1-(1+n)CF^n}{n}$ the second derivative reads
\begin{equation}
\frac{d^2H}{dt^2} = -H\Omega^2 -H^2\Biggl[\frac{3\Omega \cos \phi}{\sin \phi} W+(1+n)CF^{n-1}2p\sin \phi \Biggr]+ 2H^3 W^2
\end{equation}
At the turning points $\frac{d^2H}{dt^2}=0$  so that the first derivative has extrema
\begin{eqnarray}
\frac{dH(t_\alpha)}{dt} & = & \Bigl(\frac{dH}{dt}\Bigr)_{max}>0 \\
\nonumber
\frac{dH(t_\omega)}{dt} & = & \Bigl(\frac{dH}{dt}\Bigr)_{min}<0
\end{eqnarray}
During the expansion $\frac{dH}{dt}$ is decreasing from its positive maximum at $t_\alpha$ to its negative minimum at $t_\omega$. This corresponds to the rapid increase of positive $H$ reaching a maximum $H_{max}>0$ at $\frac{dH}{dt}=0$  followed by monotonic decrease to $H=0$ at $t_\omega$. During the contraction $\frac{dH}{dt}$ is increasing from this minimum back to the maximum at $t_{2\alpha}$. This corresponds to monotonic decrease of the negative $H$ to its minimum at $H_{min}<0$ at $\frac{dH}{dt}=0$ followed by a rapid increase to $H=0$ at $t=t_{2\alpha}$. 

During this cycle of the expansion and contraction the energy density $\bar{\rho}$ oscillates between $\bar{\rho}=0$ at $H=0$ at the two turning points and $\bar{\rho}_{max}=0.1729$x10$^{100}$ TeV/cm$^3$ at $H_{max}$ and $H_{min}$. Pressure $\bar{p}$ oscillates between $\bar{p}_\alpha=-81.0077$x10$^{100}$ TeV/cm$^3$ and $\bar{p}_\omega=+11.8525$x10$^{-10}$ TeV/cm$^3$ at the two turning points. 

%\newpage
\section{Serial acceleration - deceleration transitions in the Model A.}

\subsection{Evolution of the deceleration parameter}

The transitions from the acceleration to deceleration and vice versa of the expansion of the Universe occur at points where the deceleration parameter $q(t)=0$. In general the deceleration parameter
\begin{equation}
q(t)=\frac{-1}{H^2a}\frac{d^2a}{dt^2}=-1-\frac{1}{H^2}\frac{dH}{dt}
\end{equation}
With $\frac{dH}{dt}$ given by (6.7) we find
\begin{equation}
q(t)=-1-\frac{1}{H}\Biggl(\frac{\Omega \cos \phi}{\sin \phi}\Biggr) + \Biggl(\frac{1-(1+n)CF^n}{n}\Biggr)
\end{equation}
At small $\phi$ the expression (7.2) reads
\begin{equation}
q(\phi)=\frac{\phi^2(1-2n)-\delta^2}{2n\phi^2}
\end{equation}

Following its maximum at $t_\alpha$ the decreasing first derivative $\frac{dH}{dt}$ leads to the first transition from the accelerating to decelerating expansion. The transition occurs at small $\phi=\phi_{ad1}$ where $q(\phi_{ad1})=0$ with
\begin{equation}
\phi_{ad1}=\pm \frac{\delta}{\sqrt{1-2n}}
\end{equation}
The transition occurs at time $t_{ad1}=
\frac{\phi_{ad1}}{\Omega}=3.313461$x10$^{-38}$ s. The corresponding scale factor
\begin{equation}
a(t_{ad1})=\Bigl\{ \delta^2 +\phi_{ad1}^2 \Bigr\}^n
\end{equation}
is then equal to $a(t_{ad1})=3.883291$x10$^{-32}$. From the relation $\frac{a(t_0)}{a(t_{ad1})}=\frac{T_{ad1}}{T_0}$ we find $T_{ad1}=1.006971$x10$^{32}$ K. This corresponds to the energy of the cosmic fluid $E_{ad1}=8.677403$x10$^{15}$ TeV $\approx$ 10$^{16}$ TeV. This is the energy characteristic of the era of TOE (Theory of Everything) with a unified theory of all particle and gravity interactions.

The dependendence of the deceleration parameter on the redshift $q(z)$ in the Model A is given by (4.25) and (4.26). The fitted value $q_0=-0.2912$ of the deceleration parameter in the Model A differs substantially from the value $q_0=-0.538$ of the $\Lambda$CDM Model. In Figure 1 we show the predictions of the Model A and the $\Lambda$CDM Model for the deceleration parameter $q(z)$ for $z=0-2.34$. Below $z\sim 1$ the deceleration $q(z)$ is steeper in $\Lambda$CDM Model. The significant achievement of both Models is the prediction of the late transition from the decelerating expansion to the accelerating expansion of the Universe at the points where $q(z_{da})=0$. In $\Lambda$CDM Model this occurs at $z_{da} \approx 0.650$ ($\approx$ 5.967 Gyr ago) and in Model A at $z_{da} \approx 0.870$ ($\approx$ 7.097 Gyr ago). With the time of the deceleration-acceleration transition in Model A $t_{da}=6.713$ Gyr the corresponding scale factor $a(t_{da})=0.768717$ predicts $T_{da}=5.08687$ K, or the energy $E_{da}=0.438336$ meV. 

Figure 2 shows the time evolution of the scale factor and the deceleration parameter $q(t)$. We recognize the deceleration-acceleration transition at $t_{da}=6.713$ Gyr. It is followed by the second acceleration-deceleration transition at $t_{ad2}=38.1402$ Gyr which is a unique prediction of the Model A. A few Gyr later the deceleration parameter begins its rapid increase to its phase transition value $+\infty$ at $t=t_\omega$. The scale factor $a(t_{ad2})=4.56679$ predicts the temperature at this transition $T_{ad2}=0.856261$ K, corresponding to the energy $E_{ad2}=0.737840$x10$^{-3}$ meV. This compares with the present energy $E_0=2.34856$ meV. We anticipate low energy/temperature  quantum physics to dominate the evolution of the Cyclic Universe in this late period of its expansion.

\subsection{Evolution of the cosmic pressure $\bar{p}$ in the Friedmann Model}

During the expansion and contraction periods the pressure $\bar{p}$ changes sign. The timing of the change of sign of $\bar{p}$ is closely related to the timing of the acceleration-deceleration transitions.
It follows from the Friedmann equation (2.10) that the pressure $\bar{p}=0$ at the time $t_p$ when $\frac{dH}{dt}=-\frac{3}{2}H^2$ which corresponds to the deceleration $q(t_p)=\frac{1}{2}$. 

Using the expression (4.9) for the Hubble function $H(t)$ and the expression (6.7) for the derivative $\frac{dH}{dt}$ the condition $H^2+\frac{2}{3}\frac{dH}{dt} =0$ for $\bar{p}=0$ takes the general form
\begin{equation}
F(1-CF^n)\cos \phi = p \sin^2 \phi \Bigl[2-3n-2(1+n)CF^n \Bigr]
\end{equation}
This condition has three solutions during the expansion and the corresponding three solutions during the contraction,
\begin{table}
\caption{The evolution of deceleration parameter $q(t)$, pressure $\bar{p}$, scale factor $a(t)$, temperature $T(t)$ and the energy $E(t)$  at some significant points of time in Model A. At time $t^*$ the phase $\phi=\delta$ and $H=H_{max}$. The time $t_{60}$ marks the 60th e-folding of the initial scale factor at $t_{\alpha}$.}
\begin{tabular}{|c||c|c|c|c|c|c|}
\toprule
   & Time & $q(t)$ & $\bar{p}(t)$ & $a(t)$ & $T(t)$ & $E(t)$ \\
\colrule
$t_\alpha$ & 0 & -$\infty$ & -81.0077x10$^{100}$ TeV/cm$^3$ & 2.755601x10$^{-32}$ & 1.417x10$^{32}$ K &  1.2210x10$^{16}$ TeV \\
$t^*$ & 2.174859x10$^{-38}$ s & $<0$ & $<0$ & 3.356496x10$^{-32}$ & 1.165013x10$^{32}$ K & 1.003892x10$^{16}$ TeV \\
$t_{ad1}$ & 3.313460x10$^{-38}$ s & 0 & $<0$ & 3.883291x10$^{-32}$ & 1.006971x10$^{32}$ K & 8.677404x10$^{15}$ TeV \\
$t_{p1}$ & 5.687325x10$^{-38}$ s & $\frac{1}{2}$ & 0 & 4.951122x10$^{-32}$ & 7.897932x10$^{31}$ K & 6.805648x10$^{15}$ TeV \\
$t_{60}$ & 4.153828 yr & $>0$ & $>0$ & 3.146917x10$^{-6}$ & 1.242601x10$^{6}$ K & 107.074934 eV \\
$t_{p2}$ & 0.916785 Gyr & $\frac{1}{2}$ & 0 & 0.194743 & 20.079600 K & 1.730260 meV \\
$t_{da}$ & 6.713 Gyr & 0 & $<0$ & 0.768717 & 5.086872  K & 0.438336 meV \\
$t_0$ & 13.810 Gyr & -0.2912 & $<0$ & 1.434732 & 2.7255 K & 0.234856 meV \\
$t_{ad2}$ & 38.140174 Gyr & 0 & $<0$ & 4.566788 & 0.856261 K  & 0.737840x10$^{-6}$ eV \\
$t_{p3}$ & 42.172142 Gyr & $>0$ & 0 & 5.171192 &  0.756190 K & 0.628917x10$^{-6}$ eV \\
$t_\omega$ & 60.585933 Gyr & $+\infty$ & +11.8525x10$^{-10}$ TeV/cm$^3$ & 6.834258 & 0.572171 K & 0.493039x10$^{-6}$ eV \\ 
\botrule
\end{tabular}
\label{Table III.}
\end{table}

The first solution occurs at time $t_{p1}$ near the acceleration-deceleration transition $t_{ad1}$ at very small $\phi$ where the parameter $\delta$ is important and $F=\delta^2+\phi^2$. For $\phi$ near $\delta$ we can neglect the terms $C(\delta^2+\phi^2)^n$ in the parentheses and in the leading powers the condition (7.6) reads
\begin{equation}
\delta^2+\phi^2 = \phi^2 (2-3n)
\end{equation}
The condition for the first zero of $\bar{p}$ is then given by
\begin{equation}
\phi_{p1}=\pm \frac{\delta}{\sqrt{1-3n}}=
\pm 9.345062\text{x}10^{-56}\text{ rad}
\end{equation}
which corresponds to time $t_{p1}=\frac{\phi_{p1}}{\Omega}=5.687325$x10$^{-38}$ s. With the scale factor
\begin{equation}
a(\phi_{p1})=\delta^{2n}\Biggl(\frac{2-3n}{1-3n}\Biggr)^n=
4.951122\text{x}10^{-32}
\end{equation}
we find the temperature $T_{p1}=7.897932$x10$^{31}$ K corresponding to the energy $E_{p1}=6.805648$x10$^{15}$ TeV. The first zero of the pressure thus occurs still within the realm of TOE. With $E_{ad1}>E_{p1}$ it occurs at the beginning of the first deceleration.

An inspection of the graph of the deceleration parameter in the Figure 2 indicates two points in time where $q=\frac{1}{2}$: $t_{p2} \approx 1$ Gyr and $t_{p3} \approx 42$ Gyr. At $t_{p2}$ we can neglect $\delta$ but we can still use small $\phi$ approximation $F=2(1-\cos \phi)=\phi^2$. Then the condition (7.6) takes the form
\begin{equation}
(2-\phi^2)(1-C\phi^{2n})=(4-\phi^2)\Bigl[1-\frac{3}{2}n -(1+n)C\phi^{2n}\Bigr]
\end{equation}
Since $\phi^{2n} > \phi^2 > \phi^{2+2n}$ the solution of the equation (7.10) in the leading order reads
\begin{equation}
\phi_{p2}=\Biggl\{\frac{1-3n}{C(1+2n)}\Biggr\}^{\frac{1}{2n}}=0.047539 \text{rad}
\end{equation}
which corresponds to the time $t_{p2}=0.916785$ Gyr. The second change of sign of the pressure $\bar{p}$ thus occurs towards the end of the first deceleration period. At this time $a(t_{p2})=0.194743$ so that the temperature $T_{p2}=20.079600$ K, corresponding to the energy $E_{p2}=1.730260$ meV.

To find the third time $t_{p3}$ of the change of sign of the pressure we use $\cos \phi = \frac{1}{2}(2-F)$ to transform (7.6) in terms of $F$ as the unknown variable and solve the equation numerically for $F$ to obtain $F_{p3}=3.155512$. From this we then determine $t_{p3}=42.172142$ Gyr and the temperature $T_{p3}=0.756190$ K, or the energy $E_{p3}=6.289165$x10$^{-5}$ eV. The pressure changes sign at some 4 Gyr after the second acceleration-deceleration transition at $t_{ad2}$. 

The predictions for the deceleration parameter and the pressure during the expasion period are summarized in the Table III. The time $t_{60}$ marks the 60th e-folding of the initial scale factor at $t_{\alpha}$. In addition to the three acceleration-deceleration transitions we find three zeros in the total pressure during the expansion of the Cyclic Universe. 
The transition at $t_{da}$ has been confirmed experimentally.

\section{Evolution of Dark Energy and Dark Matter in the Model A:\\
The Generalized Friedmann Model.}

In the related paper~\cite{svec17e} we show that the Friedmann equations (2.9) and (2.10) have a general solution for the Hubble function
\begin{equation}
H^2(a)=\Bigl(\frac{a_0}{a}\bigr)^3 \Biggl\{H_0^2-\frac{8\pi G}{c^2 a_0}
\int \limits_{a_0}^a \Bigl(\frac{a'}{a_0}\bigr)^2 \bar{p} da' \Bigg\}
\end{equation}
This equation shows that the Friedmann Model is incomplete since it does not provide an indepenent information about the pressure $\bar{p}$.

We view the Universe as a thermodynamical system governed jointly by the Friedmann equations and the Laws of Thermodynamics. We thus supplement the Friedmann equations by the Euler's equation of the Thermodynamics $U=-\bar{p}V+kTS+\mu N$ where $U$ is the internal energy of the Universe, $V$ is the expanding observable volume of the Universe, $T$ is its temperature, $S$ is its total entropy and $N$ its total number of particles. $k$ is the Boltzmann constant and $\mu$ chemical potential. The First Law of Thermodynamics requires that the expression $dU=-\bar{p}dV+kTdS+\mu dN$ be fully integrable. The total differential $dU$ then splits into two parts~\cite{greiner94}
\begin{eqnarray}
dU & = & -\bar{p}dV +kTdS + \mu dN\\
0  & = & -d\bar{p}V +kdT S + d\mu N
\end{eqnarray}
The first equation is the First Law of Thermodynamics, the second equation is the Gibbs-Duhem relation~\cite{greiner94}. We assume that the Cyclic Universe is an isolated system in an equilibrium with $S=const$ and $N=const$ which satisfies the Second Law of Thermodynamics. Then the First Law reduces to the continuity equation (2.11). The Gibbs-Duhem relation gives an independent expression for the pressure
\begin{equation}
\bar{p}(t)=\bar{p}_\alpha + \int \limits_{t_\alpha}^t \frac{kdT}{dt'}\frac{S}{V(t')}dt' +\int \limits_{t_\alpha}^t \frac{d\mu}{dt'}\frac{N}{V(t')}dt'
\end{equation}

The Friedmann Hubble Function (8.1) then takes the general form~\cite{svec17e}
\begin{eqnarray}
H^2 & = & H_0^2\Bigl[\Omega_{0,0}+\Sigma_0(z)
+(1+z)^3\Bigl(\Omega_{M,0}+\Sigma_{M}(z)\Bigr)+1+z)^3\Omega_{m,0}+
+(1+z)^4\Omega_{rad,0}\Bigr]\\
   & = & H_0^2\Bigl[\tilde{\Omega}_0(z) + \tilde{\Omega}_M(z)+\tilde{\Omega}_m(z)+\tilde{\Omega}_{rad}(z)\Bigr]
\end{eqnarray}
This Generalized Friedmann Model predicts the existence of the Dark Energy $\rho_0=\frac{3c^2H_0^2}{8\pi G}\tilde{\Omega}_0(z)$ and Dark Matter $\rho_M=\frac{3c^2H_0^2}{8\pi G}\tilde{\Omega}_M(z)$ with the equations of state $w_0=-1$ and $w_M=0$, respectively, in addition to the atomic matter $\rho_m$ and radiation $\rho_{rad}$. The terms $\Omega_{0,0}$ and $\Omega_{M,0}$ are the present values of the Dark Energy and Dark Matter, respectively. The terms $\Sigma_0$ and $\Sigma_M$ are so called entropic terms since they are related to the entropies $S_0\geq 0$ and $S_M\geq 0$ carried by the Dark Energy and Dark Matter, respectively~\cite{svec17b}. They are subject to the condition
\begin{equation}
\frac{d\Sigma_0}{dz}+(1+z)^3\frac{d\Sigma_M}{dz}  =  0
\end{equation}
This condition implies $S_0(t)+S_M(t)=S=const$~\cite{svec17b}. In $\Lambda$CDM Model these terms are absent. 

With the definition of the normalized fractional energy densities $\Omega_k=\frac{8\pi G}{3 c^2 H^2} \rho_k$, $k=0,M,m,rad$ the Friedmann equations (2.9) and (2.10) take the form for all $z$ and all $t$ 
\begin{eqnarray}
\Omega_0+\Omega_{M}+\Omega_{m}+\Omega_{rad} & = & 1\\
\nonumber
w=-\Omega_0+w_m\Omega_{m}+\frac{1}{3}\Omega_{rad} & = & -\frac{1}{3}(1-2q)
\end{eqnarray}
where $w$ is equation of state in $\bar{p}=w\bar{\rho}$ and $q$ deceleration parameter. Solving (8.8) for $\Omega_0$ and $\Omega_M$ we have
\begin{eqnarray}
\Omega_0 & = & \frac{1}{3}\Bigl( 1-2q+3w_m \Omega_m +\Omega_{rad}\Bigr)\\
\nonumber
\Omega_M & = & \frac{2}{3}\Bigl(1+q -\frac{3}{2}(1+w_m)\Omega_m-2\Omega_{rad}\Bigl)
\end{eqnarray}
Since the analytical forms of $q(z)=-1+\frac{1+z}{H}\frac{dH}{dz}$ and $q(t)=-1-\frac{1}{H^2}\frac{dH}{dt}$ are known in Model A at all $z$ and all $t$, respectively, the energy densities of Dark Energy and Dark Matter can be calculated assuming the Standard Model expressions for the energy densities of  atomic matter and radiation. These are given by
\begin{eqnarray}
\Omega_m(z) & = & (1+z)^3 \Omega_{m,0}\frac{H_0^2}{H^2},   \quad
%\nonumber
\Omega_{rad}(z)  =  (1+z)^4 \Omega_{rad,0}\frac{H_0^2}{H^2}\\
\nonumber
\Omega_m(t) & = & \frac{a^3(t_0)}{a^3(t)} \Omega_{m,0}\frac{H_0^2}{H^2}, \quad
%\nonumber
\Omega_{rad}(t)  =  \frac{a^4(t_0)}{a^4(t)} \Omega_{rad,0}\frac{H_0^2}{H^2}
\end{eqnarray}
where we assume $\Omega_{m,0}=0.0484$~\cite{PDG2015} and $\Omega_{rad,0}=0.0055$~\cite{svec17b}. The predictions for the total $\Omega_0(z)$ and $\Omega_M(z)$ for $z=0-1100$ are shown in the Figures 3 and 4 and are compared with the corresponding predictions from the $\Lambda$CDM Model. The Figures show that the entropic terms are important contributions in the Model A. The predictions for the time evolution of $\Omega_0(t)$ and $\Omega_M(t)$ from 1.0098 Gyr up to 53.5176 Gyr are shown in the Figure 5. Note that at the turning points all energy densities $\rho_k$ are finite.

The notable feature of the Generalized Friedmann Model are the constraints that $\Sigma_0$ and $\Sigma_M$ vanish at $a=a_0$. These constraints determine  the present values of Dark Energy and Dark matter
\begin{equation}
\Omega_{0,0}=\Omega_0(a_0), \quad \Omega_{M,0}=\Omega_M(a_0)
\end{equation}
For Model A we obtain values
\begin{equation}
\Omega_{0,0}=0.529291, \quad \Omega_{M,0}=0.416809
\end{equation}
These values differ considerably from $\Omega_{0,0}=0.692$ and $\Omega_{M,0}=0.2596$ in the $\Lambda$CDM Model.

\begin{figure} [htp]
\includegraphics[width=12cm,height=10.5cm]{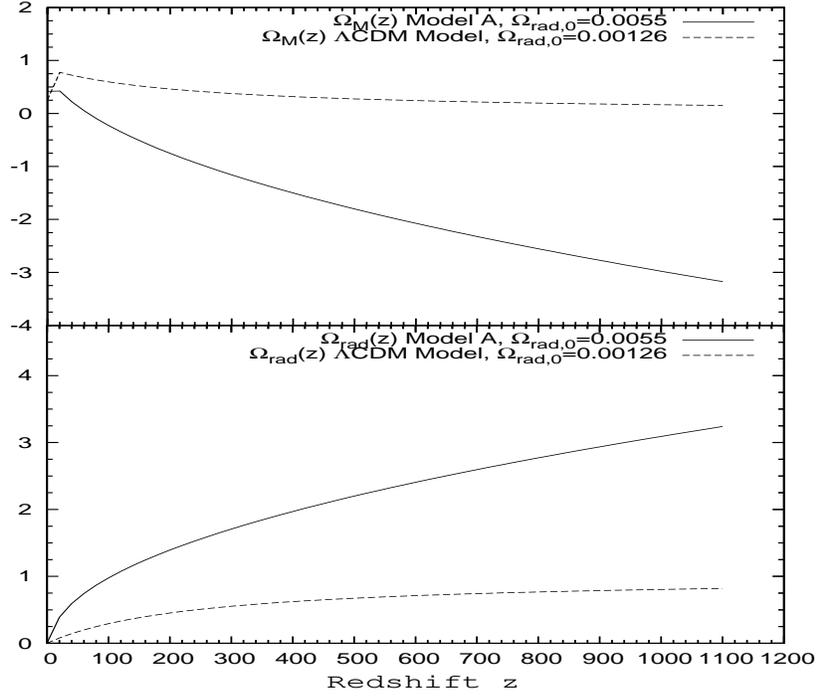}
\caption{Dark Matter (top) and radiation (bottom) at large $z$ in Model A with $w_m=0$.}
\label{Figure 3}
\end{figure}
\begin{figure} [hp]
\includegraphics[width=12cm,height=10.5cm]{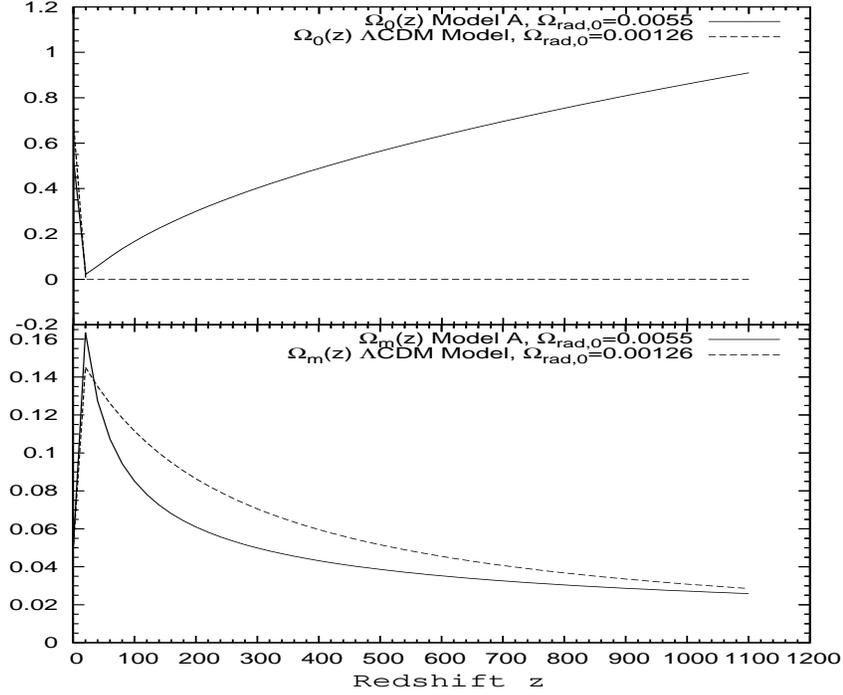}
\caption{Dark Energy (top) and atomic matter (bottom) at large $z$ in  Model A with $w_m=0$.}
\label{Figure 4}
\end{figure}

\begin{figure} [htp]
\includegraphics[width=12cm,height=10.5cm]{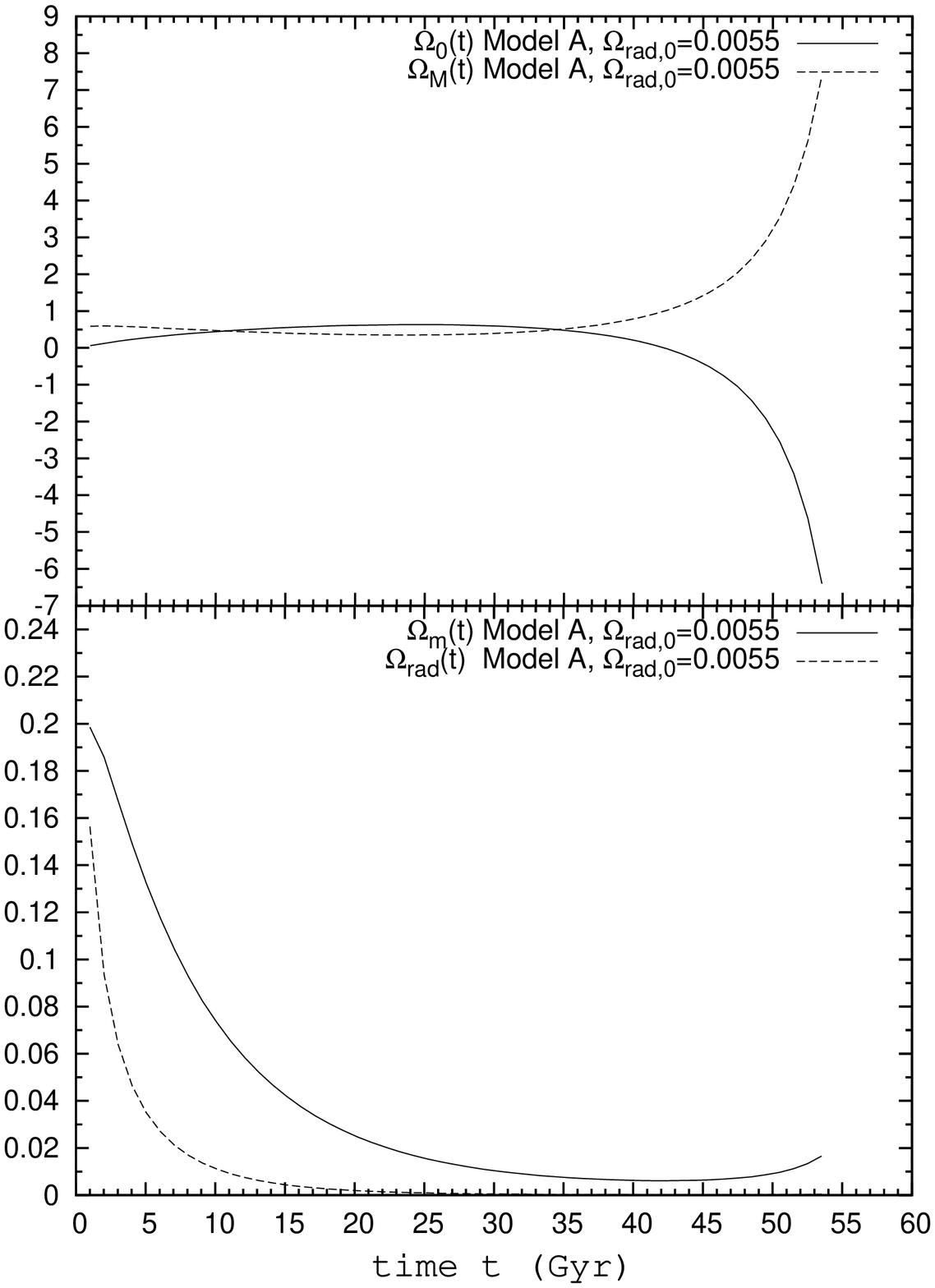}
\caption{Time evolution of Dark Energy and Dark Matter (top) and atomic matter and radiation (bottom) in Model A with $w_m=0$.}
\label{Figure 5}
\end{figure}

Using $q(t)=-1-\frac{1}{H^2}\frac{dH}{dt}$ we find that the deceleration parameter is diverging at the turning points with $q(t_\alpha)=-\infty$ and $q(t_\omega)=+\infty$. This is a classical $\lambda$ type phase transition akin e.g. to $\lambda$ phase transition of specific heat~\cite{greiner94}. In our case the order parameter is the deceleration parameter which describes the periodic phase transitions between two phases of the evolution of the Universe: the expansion and the contraction. The divergent behaviour of this order parameter should not be confused with spatial singularities. From (8.8) we find for the density $\Omega_M(t)$ \begin{eqnarray}
\begin{array} {lll}
t_\alpha \leq t <t^*, & -\infty \leq 1+q <0, & -\infty \leq \Omega_M <0\\
t=t^*, & 1+q^*=0, & \Omega_M <0\\
t^* <t \leq t^{(M)}, & 0<1+q<1+q^{(M)}, & \Omega_M <0\\
t=t^{(M)}, & 1+q^{(M)}>0, & \Omega_M =0\\
t^{(M)} <t \leq t_\omega, & 1+q^{(M)}<1+q<+\infty, & 0< \Omega_M \leq +\infty
\end{array}
\end{eqnarray}
where $1+q^{(M)}=\frac{3}{2}(1+w_m)\Omega_m(t^{(M)})+2\Omega_{rad}(t^{(M)})$. For the density $\Omega_0(t)$ we have
\begin{eqnarray}
\begin{array} {lll}
t_\alpha \leq t <t^*, & 0<1-2q \leq +\infty, & 0<\Omega_0 \leq +\infty\\
t=t^*, & 1-2q^*=3, & 0<\Omega_0\\
t^*< t<t^{(0)}, & 3>1-2q>1-2q^{(0)}, & 0<\Omega_0\\
t=t^{(0)}, & 1-2q^{(0)}<0, & \Omega_0=0\\
t^{(0)}<t\leq t_\omega, & -\infty \leq 1-2q <1-2q^{(0)}, & -\infty \leq 
\Omega_0 <0
\end{array}
\end{eqnarray}
where $1-2q^{(0)}=-3w_m \Omega_m(t^{(0)}) -\Omega_{rad}(t^{(0)})$. Since the deceleration terms cancel in $\Omega_0+\Omega_M$ there are no divergences in $H^2$. The total energy density $\bar{\rho}=0$ and the total pressure $\bar{p}$ is finite at both turning points where $H^2=0$ and $\frac{dH}{dt}$ is finite. The energy densities $\rho_k(t)$ and the pressures $p_k(t)=w_k\rho_k(t)$, $k=0,M,m,rad$ are also all finite at both turning points. The negative energy densities suggest that the components form a bound system.

Assuming that the Hubble function $H^2$ in (8.5) is identified with the Hubble function of the Model A, $H^2=H^2_A$, we can determine the entropic terms $\Sigma_0$ and $\Sigma_M$ in the Model A. Taking a derivative of (8.5) by $da$ we find using (8.7)
\begin{equation}
\Sigma_M(a)=-\frac{1}{3}\Bigl(\frac{a}{a_0}\Bigr)^3\frac{a}{H_0^2}\frac{dH^2_A}{da}-\frac{4}{3}\Bigl(\frac{a_0}{a}\Bigr) \Omega_{rad,0}-\Omega_{M,0}-\Omega_{m,0}
\end{equation}
where $H_0=H_{A,0}=67.81$ kms$^{-1}$Mpc. Using the relation (8.7) and integrating it over $da$ we find
\begin{equation}
\Sigma_0(a)= \frac{1}{H_0^2}[H_A^2-H_{A,0}^2]+
\Bigl[1-\Bigl(\frac{a_0}{a}\Bigr)^3 \Bigr] 
[\Omega_{M,0}+\Omega_{m,0}+\Sigma_M]  + 
\Bigl[1-\Bigl(\frac{a_0}{a}\Bigr)^4 \Bigr]\Omega_{rad,0}
\end{equation}
With the expression for $H_A$ given by (4.17) we find
\begin{equation}
\frac{dH^2_A}{da}=-\frac{2\Omega^2n(1+aC)}{a^{\frac{1+n}{n}}}
\Bigl[naC\Bigl(a^{\frac{1}{n}}-(1+aC)^{\frac{1}{n}}\Bigr)+2(1+aC)^{\frac{1}{n}}\Bigr]
\end{equation}
where $\Omega=0.0518535$ radGyr$^{-1}=0.0518535\times$997.78 kms$^{-1}$Mpc. At $a=a_0$ the condition $\Sigma_M(a_0)=0$ in (8.15) implies a condition for $\Omega_{M,0}$
\begin{equation}
-\frac{a_0}{H^2_0}\frac{dH^2_A}{da}_{|_0} = 3\Omega_{M,0}+3\Omega_{m,0} +4\Omega_{rad,0}
\end{equation}
Then $\Omega_{0,0}=1-\Omega_{M,0}-\Omega_{m,0} -\Omega_{rad,0}$. With $\Omega_{m,0}=0.0484$ and $\Omega_{rad,0}=0.0055$ we recover exactly the values (8.12) for $\Omega_{M,0}$ and $\Omega_{0,0}$ demonstrating the validity of the assumption $H^2=H^2_A$. 

\section{The spatial curvature and its evolution in the Model A}

Using the equation (2.4) we define a fractional curvature energy density
\begin{equation}
\Omega_c=\frac{8\pi G}{3c^2 H^2}\rho_c = \frac{-kc^2}{R_0^2a^2H^2}
=\Biggl\{\frac{-kc^2}{R^2_0a_0^2}\frac{(1+z)^2}{H_0^2}\Biggr\}
\frac{H_0^2}{H^2(z)}
=\Bigl\{\Omega_{c,0}(1+z)^2\Bigr\}
\frac{H_0^2}{H^2(z)}
\end{equation}
In a related paper~\cite{svec17b} we fit the Model A to the angular diameter distance data and determine the present value of the curvature density $\Omega_{c,0}=0.831943$ and the curvature parameter $R_0=11.0187$ Glyr. The positive value of $\Omega_{c,0}$ corresponds to the negative spatial curvature $k=-1$ indicating an open infinite anti-de Sitter spacetime. The Etherington relation between angular diameter distance data and luminosity distance data was modified by a correction factor $\eta(z)=\frac{1}{1+\eta_0z}$ with $\eta_0=0.0150$ to obtain $\Omega_{c,0}=0.004000$ for $\Lambda$CDM Model. Our value for $\eta_0$ is in excellent agreement with the very recent model independent test of the similarly corrected Etherington relation which yields $\eta_0=0.0147^{+0.056}_{-0.066}$~\cite{ruan18}.

The proper volume of the Universe is given by $V(t)=a^3(t)\mathcal{V}$ where $\mathcal{V}$ is the comoving volume. Irrespective of its shape we assume that $\mathcal{V}=R_0^3$ where $R_0$ is the curvature parameter in the Robertson-Walker metric. We define a proper extent of the Universe $R(t)=a(t)R_0$ and the velocity of its expansion $\frac{dR}{dt}=H(t)R(t)$. Evolution of the proper extent and its velocity during the expansion period at selected times are presented in the Table IV together with the Hubble function. It follows from (9.1) that the  evolution of $\Omega_c$ is also given by
\begin{equation}
\Omega_c(t)=\frac{-kc^2}{H(t)^2R(t)^2}=\frac{-kc^2}{(\frac{dR}{dt})^2}
\end{equation}

In our terminology the term "inflation" describes the incredibly fast rise of the Hubble function from $H=0$ at $t_\alpha$ to $H_{max}=4.1294x10^{53}$ Gyr$^{-1}$ during the first 2.174859x10$^{-38}$ s. "Pre-inflation" describes the equally fast rise from $H_{min}=-4.1294x10^{53}$ Gyr$^{-1}$ to $H=0$ during the same period of time. During this "inflation" there is only a small change in the scale factor. In contrast, the Standard Model Inflation refers to an exponential rise of the scale factor $a(t)$. The Standard Model Inflation begins at the onset of GUT (Grand Unified Theory) at $E_{GUT}=10^{12}$ TeV and lasts about $\sim 10^{-34}$ s with some 60 e-foldings of the scale factor during this time. In Model A the 60 e-foldings of the initial scale factor $a(t_\alpha)$ take more than 4 years. 

We conclude, that there is no Standard Model Inflation in the Model A. Instead there is a very rapid Hubble function "inflation" followed by a rapid change of the sign of the pressure. This is a unique prediction of the Model A that is testable in astronomical observations. 

\begin{table}
\caption{The evolution of the proper extent $R(t)$ of the Cyclic Universe, Hubble function $H(t)$, the velocity of the expansion $\frac{dR}{dt}$ and the fractional curvature density $\Omega_c(t)$ at some significant points of time in the Model A.}
\begin{tabular}{|c||c|c|c|c|c|c|c|c|}
\toprule
Time & $t_\alpha$ & $t^*$ & $t_{ad1}$ & $t_{60}$ & $t_{da}$ & $t_0$ & $t_{ad2}$ & $t_\omega$ \\
\colrule
$R(t)$ (Glyr) & 2.8745 $\mu m$ & 3.5014 $\mu m$ & 4.0509 $\mu m$ & 3.4958x10$^{-5}$ & 8.4703 & 15.8088 & 50.3200 & 75.3047 \\
$H(t)$ (Gyr$^{-1}$) & 0 & 4.1294x10$^{53}$ & 3.7887x10$^{53}$ & 1.3703x10$^{8}$ & 0.11797 & 0.069351 & 0.003310 & 0 \\
$\frac{dR}{dt}$ (c) & 0 & 1.5295x10$^{23}$c & 1.6211x10$^{23}$c & 4.7513x10$^3$c & 0.99895c & 1.0964c & 0.16660c & 0 \\
$\Omega_c(t)$ & $\infty$ & 4.2747x10$^{-47}$ & 3.8052x10$^{-47}$ & 4.4297x10$^{-8}$ & 1.0021 & 0.831943 & 36.0288 & $\infty$ \\
\botrule
\end{tabular}
\label{Table IV.}
\end{table}

In the Standard $\Lambda$CDM Model the curvature $\Omega_c$ is constant and $\Omega_c=0$. In the Model A the curvature $\Omega_c(t)$ is dynamic and evolving with the expansion of the Universe. The Table IV shows the values of $\Omega_c(t)$ at some significant points in time. We note that while $\Omega_c$ diverges at the turning points, the energy density $\rho_c$ is finite. The Table IV shows that the curvature in the Model A evolves from a clearly flat early Universe to increasingly curved late Universe. Importantly, at z=1100 (t=0.000164) corresponding to the Cosmic Microwave Background the  Universe was still very flat with $\Omega_c(CMB)$=8.4250x10$^{-10}$. It is interesting to note that in the Model A the rate of change of $\Omega_c$ is given by 
\begin{equation}
\frac{d\Omega_c}{dt}=2q(t)H(t)\Omega_c(t)
\end{equation}
At the extrema the deceleration parameter $q(t)=0$ and
\begin{equation}
\frac{d^2\Omega_c}{dt^2}=2H\Omega_c\frac{dq}{dt}
\end{equation}
so that the kind of the extremum of $\Omega_c$ is controlled by the sign of $\frac{dq}{dt}$. Thus $\Omega_c$ has a local minimum at $t_{ad1}$, a local maximum at $t_{da}$ and another local minimum at $t_{ad2}$.  

The principal motivation for the Inflation Model, which is a part of the Standard Model, was to resolve the "flatness problem" by demonstrating the flatness of the early Universe. The Model A naturally exhibits flat early Universe and the apparent flatness of the present Universe at scales up to several 100 Mpc. A negative spatial curvature $\Omega_c(t)$ evolving from the initially flat spacetime of $\Lambda$CDM Model emerges naturally in numerical General Relativity~\cite{bolejko17}. It peaks at $t \sim 10$ Gyr with $\Omega_c \sim $ 0.10 and slowly decreases below $t=13.8$ Gyr. A flat spacetime at the early Universe does not necesserily imply a constant flat spatial curvature $\Omega_c(t)=0$.

In general, the determination of the curvature $\Omega_{c,0}$ from the  angular diameter distance data depends on the assumed Hubble function $H(z)$ and on the opacity parameter $\epsilon$~\cite{svec17b,wang17} which quantifies a violation of the Etherington relation~\cite{svec17b,wang17} which relates the angular diameter distance $d_A(z)$ to the luminosity distance $d_L(z)$. It is interesting to compare our values of $\Omega_{c,0}$ for $\Lambda$CDM Model and Model A with the recent model independent determination of $\Omega_{c,0}$~\cite{wang17}. Assuming a reconstructed Hubble function $H(z)$ these authors find that the best fit of $\Omega_{c,0}$ correlates with the value and the error of $H_0$ as well as the opacity $\epsilon$. The results of these analyses are shown in the Table V. For $H_0=73.24 \pm 1.74$ kms$^{-1}$Mpc and $\epsilon=-0.232\pm 0.075$ they find $\Omega_{c,0}=1.044\pm 0.514$ which is consistent with our result  $\Omega_{c,0}=0.831943\pm 0.239578$ in the model independent Model A with $H_0=67.81\pm 0.92$ kms$^{-1}$Mpc and opacity $\epsilon=-0.015$~\cite{svec17b}.

\begin{table}
\caption{Comparison of the present spatial curvature $\Omega_{c,0}$ in $\Lambda$CDM Model and Model A from our data analysis in Ref.~\cite{svec17b} and the three cases (a),(b),(c) from the data analysis in Ref.~\cite{wang17}.}
\begin{tabular}{|c||c|c|c|c|c|}
\toprule
Model & $\Lambda$CDM & Model A & Case (a) & Case (b) & Case (c) \\
\colrule
$H_0$ [kms$^{-1}$Mpc] & 67.81$\pm$0.92 & 67.81$\pm$0.92 & 67.56$\pm$4.77 & 67.74$\pm$0.46 & 73.24$\pm$1.74 \\
$\epsilon$ & -0.015 & -0.015 & -0.018$\pm$0.084 & -0.012$\pm$0.075 & -0.232$\pm$0.075 \\
$\Omega_{c,0}$ & 0.004$\pm$0.266 & 0.832$\pm$0.240 & 0.440$\pm$0.645 & 0.374$\pm$0.580 & 1.044$\pm$0.514 \\
\botrule
\end{tabular}
\label{Table V.}
\end{table}

%\newpage
\section{Conclusions and Outlook.}

We have constructed two analytical models of the scale factor $a(t)$ based on the requirements of "general solvability" and periodicity. There are no initial or final state spacetime singularities. Consequently, there is no breakdown of the Einstein theory of gravity at the turning points. The non-zero value of the initial scale factor $a(t_\alpha)$ is given by the Planck temperature. The two scale factors predict two models called Model A and Model C for the Hubble functions $H(t)$ and $H(z)$. In both Models the Hubble functions are distinguished by their zero values at the turning points. The fits of the Hubble data select the Model A as the Cyclic Universe. The Cyclic Universe is Eternal.  

In the absence of Dark Energy and Dark Matter in the Hubble function the Friedmann equations predict only a deceleration of the expansion. With the presence of a constant Dark Energy $\Omega_{0,0}=\Omega_\Lambda$ and a constant Dark Matter $\Omega_{M,0}=\Omega_M$ in the Hubble function $H^2$, the $\Lambda$CDM Model predicts only one deceleration-acceleration transition at late times $\approx$ 5.967 Gyr ago. The scale factor in the Model A predicts a similar transition $\approx$ 7.097 Gyr ago. There are two additional transitions at $t_{ad1}$ and $t_{ad2}$. In addition to the three acceleration-deceleration transitions we find three zeros in the total pressure during the expansion of the Cyclic Universe. These findings are a unique feature of the Model A which imply dynamical Dark Energy and Dark Matter with additional entropic terms $\Sigma_0$ and $\Sigma_M$ in the Hubble function $H^2$. The dependence of the entropic Dark Energy and Dark Matter terms on the redshift $z$ or cosmic time $t$ in the Model A is fully determined by the Friedmann equations and arises from the cyclicity of the non-singular scale factor that defines the input deceleration parameter. Friedmann equations are the Einstein equations for the Robertson-Walker gravitational field of the homogeneous and isotropic spacetime. 

With no dynamical terms of Dark Energy and Dark Matter, the Standard $\Lambda$CDM Model predicts a flat Universe with constant curvature density $\Omega_c=0$. The scale factor in the Model A predicts a dynamic curvature density $\Omega_c(z)$ evolving from a flat early Universe to a curved anti-de Sitter Universe at the present time. This evolution of $\Omega_c(z)$ is thus connected to the presence of the entropic terms $\Sigma_0$ and $\Sigma_M$ in the Hubble function $H^2$ of the Model A.

We find that there is no Standard Model Inflation in the Model A. Instead, near the initial turning point, there is a very rapid increase ("inflation") of the Hubble function followed by a rapid change of the sign of the pressure. This is a unique prediction of the Model A that is open to tests in astronomical observations.

While the rapid increase of the Hubble function physically relates to the rapid   increase of the energy density $\bar{\rho}$ and the rapid change in the pressure $\bar{p}$ at $t>0$ near $t_\alpha$, the velocity $\frac{dR}{dt}$ physically relates to the rate of change of the spatial extent of the Universe. The notable feature of this velocity of the expansion of the Universe are its extremally high superluminal values during the periods of TOE and GUT at Planck energies $10^{16}-10^{10}$ TeV. Such unusual expanding flat spacetimes may not be compatible with the physics of subluminal baryonic particles and fields considered so far in the Minkowski spacetime of the Standard Model of the particle physics. Perhaps the unification of all forces and gravity expected at TOE and GUT energies may require a new physics involving tachyonic fields and particles. 

LIGO detections of gravity waves GW150914~\cite{abbott16a}, GW151226~\cite{abbott16b} and GW170104~\cite{abbott17a} in binary black hole mergers constrain the the gravitational waves propagation speed $c_{gr}$ to a broad interval $0.55c < c_{gr}<1.42c$ with $90\%$ confidence level~\cite{cornish17}. Recent detection of GW170817 from a binary neutron star inspiral by LIGO-VIRGO~\cite{abbott17c} in association with a detection of a gamma ray burst GRB170817A from the same event by Fermi-GBM~\cite{goldstein17} allowed to make an extraordinarily precice measurement of the speed of gravitational waves $c_{gr}$. It is equal to the speed of light within $\sim$ 1:10$^{-15}$~\cite{abbott17d}. This finding exludes a number of gravity theories that require $c_{gr} < c$ or $c_{gr} > c$. However this finding does not necesserily exclude unified theories TOE and GUT with new tachyon particles and fields. 

Since the tail of the superluminal velocities of the expansion persists to the time $t_{60}=4.1538$ yr and energies $\sim 107$ eV it may leave some imprint on the particle and nuclear physics of this time. While we observe today hadron resonances (peaks) at resonant masses $m_R^2>0$ produced in particle scattering in energy variables $s>0$, the possibility of finding tachyon resonant structures (dips) in momentum transfer variables $t<0$ with $m_T^2<0$ in particle scattering processes with polarized targets is still entirely unexplored. Such laboratory findings of tachyons could provide a valuable connection to the physics of the very early stages of the expansion of the Universe. 

In the Model A the entire evolution of the Cyclic Universe is described by an energy function - the Hubble function $H(t)$. The evolution proceeds by the minimization of this energy function from its maximum $+H_{max}>0$ to its minimum $-H_{max}<0$ at times $t^*$=2.175x10$^{-38}$s$>0$ and $T-t^*$, respectively, followed by a rapid phase transition from $-H_{max}<0$ to $+H_{max}>0$. There is no Big Bang singularity at $t_\alpha$ or $t_{2\alpha}=T$ where $H(t_\alpha)=H(t_{2\alpha})=0$. This cyclic evolution dynamics of energy function minimization may represent a new general evolution principle of the Universe.

The known analytical form of the Hubble functions $H(z)$ and $H(t)$ in the Model A enables to predict from the Friedmann equations the analytical form of the energy densitities of the Dark Energy and Dark Matter at any $z$ or any $t$. The Model A predicts the emergence of an evolving negative spatial curvature from a deeply flat spacetime at the early Universe. These predictions make the Model A testable in the observations to be made by the ongoing and upcoming astronomical surveys at high redshifts including Dark Energy Survey (DES)~\cite{DES}, Large Synoptic Survey Telescope (LSST)~\cite{LSST}, Euclid Mission~\cite{Euclid}, Wide Field Infrared Survey Telescope (WFIRST)~\cite{WFIRST} and Square Kilometer Array (SKA)~\cite{SKA}. The Model A can also serve as a useful theoretical tool to explore the physics of the evolution of the Universe in the past at very high $z$ and throughout its future.

\acknowledgements

I acknowledge with thanks the technical support of this research by Physics Department, McGill University.

\newpage
\appendix

\section{Table of the Hubble data $H(z)$ and $AH(z)$}

\begin{table} 
\caption{Measured Hubble parameter data $H(z)$ and the predicted Hubble parameter $AH(z)$ from the best fit A.01 of the Model A to the Hubble data.}
\begin{tabular}{|c||c|c|c|c|c|}
\toprule 
$z$ & $H(z)$ & $\sigma_H$ & $AH(z)$ & $A\sigma_{H}$ & Reference \\
\colrule
    & ($kms^{-1}Mpc^{-1}$) & ($kms^{-1}Mpc^{-1}$) & ($kms^{-1}Mpc^{-1}$) & ($kms^{-1}Mpc^{-1}$) &    \\
\colrule    
0.000 & 67.81 &  0.92 & 67.81 & 0.92 & Planck 2015~\cite{planck15}\\
0.070 & 69.00 & 19.60 & 71.21 & 0.42 & Zhang {\sl et al.}~\cite{zhang14}\\
0.090 & 69.00 & 12.00 & 72.19 & 0.54 & Jimenez {\sl et al.}~\cite{jimenez03}\\
0.120 & 68.60 & 26.20 & 73.68 & 0.71 & Zhang {\sl et al.}~\cite{zhang14}\\
0.170 & 83.00 & 8.00 & 76.20 & 0.98 & Simon {\sl et al.}~\cite{simon05}\\
0.179 & 75.00 & 4.00 & 76.65 & 1.02 & Moresco {\sl et al.}~\cite{moresco12}\\
0.199 & 75.00 & 5.00 & 77.67 & 1.13 & Moresco {\sl et al.}~\cite{moresco12}\\
0.200 & 72.90 & 29.60 & 77.72 & 1.13 & Zhang {\sl et al.}~\cite{zhang14}\\
0.240 & 79.69 & 3.32 & 79.78 & 1.33 & Gaztanaga {\sl et al.}~\cite{gaztanaga09}\\
0.270 & 77.00 & 14.00 & 81.34 & 1.47 & Simon {\sl et al.}~\cite{simon05}\\
0.280 & 88.80 & 36.60 & 81.86 & 1.52 & Zhang {\sl et al.}~\cite{zhang14}\\
0.352 & 83.00 & 14.00 & 85.68 & 1,84 & Moresco {\sl et al.}~\cite{moresco12}\\
0.400 & 95.00 & 17.00 & 88.27 & 2.03 & Simon {\sl et al.}~\cite{simon05}\\
0.430 & 91.80 & 5.30 & 89.90 & 2.15 & Moresco {\sl et al.}~\cite{moresco16}\\
0.480 & 97.00 & 62.00 & 92.66 & 2.33 & Stern {\sl et al.}~\cite{stern10}\\
0.593 & 104.00 & 13.00 & 99.05 & 2.70 & Moresco {\sl et al.}~\cite{moresco12}\\
0.680 & 92.00 & 8.00 & 104.11 & 2.95 & Moresco {\sl et al.}~\cite{moresco12}\\
0.781 & 105.00 & 12.00 & 110.13 & 3.18 & Moresco {\sl et al.}~\cite{moresco12}\\
0.875 & 125.00 & 17.00 & 115.89 & 3.36 & Moresco {\sl et al.}~\cite{moresco12}\\
0.880 & 90.00 & 40.00 & 116.20 & 3.37 & Stern {\sl et al.}~\cite{stern10}\\
0.900 & 117.00 & 23.00 & 117.44 & 3.41 & Simon {\sl et al.}~\cite{simon05}\\
1.037 & 154.00 & 20.00 & 126.13 & 3.59 & Moresco {\sl et al.}~\cite{moresco12}\\
1.300 & 168.00 & 17.00 & 143.62 & 3.79 & Simon {\sl et al.}~\cite{simon05}\\
1.363 & 160.00 & 33.00 & 147.97 & 3.82 & Moresco~\cite{moresco15}\\
1.430 & 177.00 & 18.00 & 152.66 & 3.85 & Simon {\sl et al.}~\cite{simon05}\\
1.530 & 140.00 & 14.00 & 159.79 & 3.91 & Simon {\sl et al.}~\cite{simon05}\\
1.750 & 202.00 & 40.00 & 176.00 & 4.15 & Simon {\sl et al.}~\cite{simon05}\\
1.965 & 186.50 & 50.40 & 192.53 & 4.70 & Moresco~\cite{moresco15}\\
2.340 & 222.00 & 7.00 & 222.98 & 6.78 & Delubac {\sl et al.}~\cite{delubac15}\\
\botrule
\end{tabular}
\label{Table VI.}
\end{table}

\end{document}